\documentclass[pre,twocolumn,superscriptaddress]{revtex4}

\usepackage{amsmath,amssymb}
\usepackage{graphicx}
\usepackage{bm}
\usepackage{algorithmicx}
\usepackage{algorithm}
\usepackage[noend]{algpseudocode}
\usepackage{color}
\usepackage[colorlinks,linkcolor=blue,anchorcolor=blue,citecolor=blue]{hyperref}
\usepackage[T1]{fontenc}

\begin{document}

\title{Bayesian deep learning for error estimation in the analysis of
anomalous diffusion}

\author{Henrik Seckler}
\affiliation{Institute for Physics \& Astronomy, University of Potsdam, 14476 Potsdam-Golm, Germany}
\author{Ralf Metzler}\email{rmetzler@uni-potsdam.de}
\affiliation{Institute for Physics \& Astronomy, University of Potsdam, 14476 Potsdam-Golm, Germany}
\affiliation{{Asia Pacific Centre for Theoretical Physics, Pohang 37673,
Republic of Korea}}

\date{\today}

\begin{abstract}
Modern single-particle-tracking techniques produce extensive time-series of
diffusive motion in a wide variety of systems, from single-molecule motion
in living-cells to movement ecology. The quest is to decipher the physical
mechanisms encoded in the data and thus to better understand the probed
systems. We here augment recently proposed machine-learning techniques
for decoding anomalous-diffusion data to include an uncertainty estimate
in addition to the predicted output. To avoid the Black-Box-Problem a
Bayesian-Deep-Learning technique named Stochastic-Weight-Averaging-Gaussian
is used to train models for both the classification of the diffusion
model and the regression of the anomalous diffusion exponent of
single-particle-trajectories. Evaluating their performance, we find that
these models can achieve a well-calibrated error estimate while maintaining
high prediction accuracies. In the analysis of the output uncertainty
predictions we relate these to properties of the underlying diffusion models,
thus providing insights into the learning process of the machine and the
relevance of the output.
\end{abstract}

\maketitle

\section{Introduction}
\label{sec_intro}

In 1905 Karl Pearson introduced {the concept of the random walk as
a path of successive random steps} \cite{pearson_1905}. The model has since
been used to describe random motion in many scientific fields, including
ecology \cite{okubo_1986, vilk_origins}, psychology \cite{psychology},
physics \cite{physics}, chemistry \cite{chemistry}, biology \cite{biology}
and economics \cite{malkiel_wallstreet,bouchaud_finance}.  As long as the
increments (steps) of such a random walk are independent and identically
distributed with a finite variance, it will, under the \textit{Central Limit
Theorem} (CLT) \cite{central_limit_theorem}, lead to normal diffusion {in the
limit of many steps}. The prime example of this is \textit{Brownian motion},
which describes the random motion of small particles suspended in liquids or
gases \cite{einstein_brown,smoluchowski_brown,sutherland_brown,langevin_brown}.
Amongst others, normal diffusion entails that the \textit{mean squared
displacement} (MSD) grows linearly in time \cite{kampen_stochproc,
levy_processus, hughes_randomwalks}, $\langle \mathbf{r}^2(t)\rangle \propto
K_1 t$.

\begin{figure}
\includegraphics[width=\linewidth]{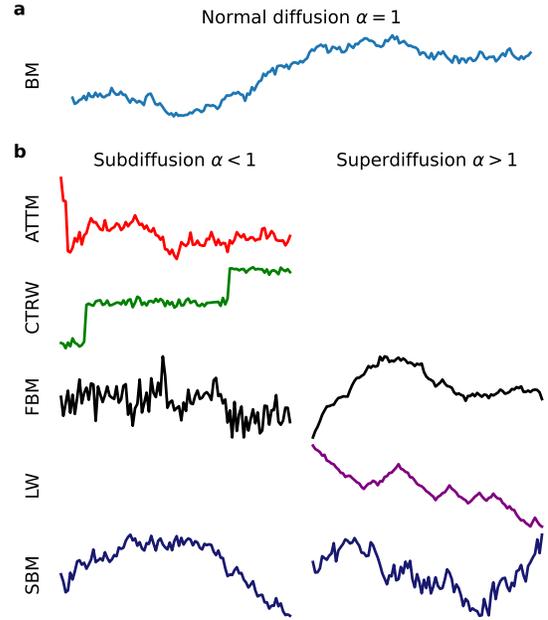}
\caption{Sample trajectories of normal (a) and anomalous (b) diffusion. All
shown trajectories are corrupted by white Gaussian noise.}
\label{fig_trajplot}
\end{figure}

In practice however many systems instead exhibit a power law behaviour
$\langle \mathbf{r}^2(t)\rangle\propto K_\alpha t^\alpha$ of the
MSD \cite{bouchaud_anodif,metzler_guide,golding_bact_subdif,
manzo_ergodicity, krapf_spectral,
stadler_nonequ,kindermann_nonergodic,sokolov,hoefling_crowded,horton_crowded,norrelykke,leijnse,saxton1,saxton2,burov,ernst},
indicating that one or several conditions of the CLT are not fulfilled. We
refer to such systems as \textit{anomalous diffusion}. A motion with anomalous
diffusion exponent $0<\alpha<1$ is called subdiffusive, whereas for $\alpha>1$
it is referred to as superdiffusive {(including ballistic motion with
$\alpha=2$)}. In order to describe such systems mathematically, many models
have been proposed, in which one or multiple conditions of the CLT are broken
\cite{bouchaud_anodif,metzler_guide,metzler_models}. {Some important examples
(see section {\ref{sec_andi}} for details) of such models are continuous-time
random walk (CTRW) {\cite{montroll_ctrw,hughes_ctrw,weissmann_ctrw}},
fractional Brownian motion (FBM) {\cite{mandelbrot_fbm}}, L\'evy walk
(LW) {\cite{levy_flight, chechkin_levyflight, shlesinger_levywalk,
zaburdaev_levywalk}}, scaled Brownian motion (SBM) {\cite{lim_sbm, jeon_sbm}}
and annealed transient time motion (ATTM) {\cite{massignan_attm}}.} Sample
trajectories for these are shown in figure \ref{fig_trajplot}.

\begin{figure*}
\fbox{\includegraphics[width=0.98\linewidth]{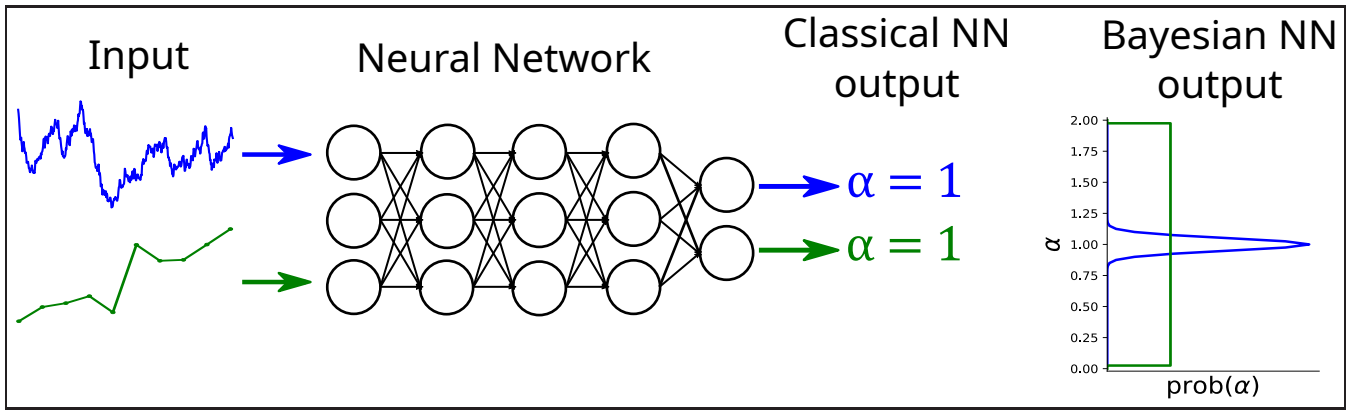}}
\caption{Illustration of the problem of reliability in deep learning. The
illustration depicts the use of a neural network to predict the anomalous
exponent $\alpha$ for two sample trajectories. Despite receiving severely
different inputs, a classical neural network may still predict the same output
(anomalous diffusion exponent $\alpha=1$) for both cases. The difference
between the outputs only becomes clear when predicting not just the output
itself but a distribution over all possible outputs, as it is done, for
example, in Bayesian Deep Learning.}
\label{fig_intro}
\end{figure*}

As each of these models correspond to different sources of anomalous
diffusion, determining the model underlying given data can yield useful
insights into the physical properties of a system \cite{meroz_toolbox,
cherstvy_mucin, golding_bact_subdif, manzo_ergodicity, krapf_spectral,
stadler_nonequ,kindermann_nonergodic}. Additionally one may wish to determine
the parameters attributed to these models, the most sought-after being the
anomalous diffusion exponent $\alpha$ {and the generalised diffusion coefficient
$K_\alpha$} \cite{makarava_bayeshurst, golding_bact_subdif}. The used
experimental data typically consist of single particle trajectories,
such as the diffusion of a molecule inside a cell \cite{elf_moleculedif,
cherstvy_mucin, hoefling_crowded,horton_crowded, norrelykke,leijnse, biology},
the path of an animal \cite{okubo_1986, vilk_origins, bartumeus_animotion}
or the movement of stock prices \cite{malkiel_wallstreet, plerou_stockprice}.

Plenty of techniques have been developed to tackle these tasks,
usually through the use of statistical observables. {Some examples
include the ensemble averaged or time averaged MSD to determine the
anomalous diffusion exponent and/or differentiate between a non-ergodic
and ergodic model {\cite{metzler_singlepartana}}, the p-variation test
{\cite{magdziarz_pvari}}, the velocity auto correlation for differentiation
between CTRW and FBM {\cite{burov}}, the single trajectory power spectral
density to determine the anomalous diffusion exponent and differentiate
between models {\cite{metzler_bmbeyond,vilk_animal_psd}}, the first
passage statistics {\cite{condamin_firstpassage}} and the codifference
{\cite{slezak_codifference}}.} Such techniques may struggle when the
amount of data is sparse and, with its rise in popularity, successful
new methods using machine learning have emerged in recent years
\cite{granik,munozgil_machinelearn,pinholt_fingerprint}.

In an effort to generalise and compare the different approaches
the \textit{Anomalous Diffusion (AnDi) Challenge} was held in 2020
\cite{munozgil_andich_start,munozgil_andich_end}. The challenge
consisted of three tasks, among them the determination of the
anomalous diffusion exponent $\alpha$ and the underlying diffusion
model from single particle trajectories. The entries included a wide
variety of methods ranging from mathematical analysis of trajectory
features \cite{aghion, meyer}, to Bayesian Inference \cite{krog_bay1,
park_bay2,thapa_bay3}, to a wide variety of machine learning techniques
\cite{li,gentili,argun,verdier,manzo_elm,granik,garibo,janczura,kowalek,
loch-olszewska,bo}. While the best results were achieved by deep learning
(neural networks), this approach suffers from the so-called \textit{Black
Box Problem}, delivering answers without providing explanations as to how
these are obtained or how reliable they are \cite{szegedy_black_box_problem}.
In particular, outputs are generated even in situations when the neural network
was not trained for the specific type of motion displayed by the system under
investigation.  In this work, we aim at alleviating this problem by expanding
the deep learning solutions to include an estimate of uncertainty in the
given answer, as illustrated in figure \ref{fig_intro}. This is a feature
that other techniques like \textit{Bayesian Inference} can intrinsically
provide \cite{krog_bay1, park_bay2,thapa_bay3}.

Such a reliability estimation is a well known problem in machine learning. For
neural networks the solutions vary from the calibration of neural network
classifiers \cite{degroot_forecast,guo_reliability,naeini_ece,levi_ence},
to using an ensemble of neural networks and obtaining an uncertainty from
the prediction spread \cite{lakshminarayanan_deep_ensemble}, to fully
modelling the probability distribution of the outputs in \textit{Bayesian
Neural Networks} \cite{mackay_bnn}. In recent years the latter has been
expanded to be applicable to deep neural networks without resulting in
unattainable computational costs. These \textit{Bayesian Deep Learning}
(BDL) techniques approximate the probability distribution by various means,
for instance, by using drop out \cite{gal_dropout, gal_phdthesis} or an
ensemble of neural networks \cite{lakshminarayanan_deep_ensemble}. We here
decided on using a method by Maddox et al. named \textit{Stochastic Weight
Averaging Gaussian} (SWAG), in which the probability distribution over the
network weights is approximated by a Gaussian, obtained by interpreting a
stochastic gradient descent as an approximate \textit{Bayesian Inference}
scheme \cite{maddox_swag,wilson_multiswag}.  We find that these methods are
able to produce well calibrated uncertainty estimates, while maintaining the
prediction performance of the best \emph{AnDi-Challenge} solutions. We show
that analysing these uncertainty estimates and relating them to properties
of the diffusion models can provide interesting insights into the learning
process of the machine.

The paper is structured as follows. A detailed analysis
of our results for regression and classification is presented in section
\ref{sec_main}. These results are then discussed and put into perspective
in section \ref{sec_last}. A detailed explanation of the utilised methods is
provided in section \ref{sec_mainmethods}. Here we provide a brief introduction
to the different anomalous diffusion models in subsection \ref{sec_andi}
and the used SWAG method in subsection \ref{sec_bdl}. Subsequently the neural
network architecture and training procedure used in our analysis is presented
in subsection \ref{sec_archi}. The Supplementary Information details the
reliability assessment methods and provides supplementary figures.

\section{Results}
\label{sec_main}

In the following we employ the Methods detailed in
section \ref{sec_mainmethods} to construct the \textit{Multi-SWAG}
\cite{wilson_multiswag} models and use these to determine the anomalous
diffusion exponent $\alpha$ or the diffusion model of computer generated
trajectories. We also provide detailed error estimates to qualify the given
outputs. These estimates consist of a standard deviation for regression
and model probabilities for classification. The trajectories are randomly
generated from one of the five diffusion models: continuous-time random walk
(CTRW) \cite{montroll_ctrw,hughes_ctrw,weissmann_ctrw}, fractional Brownian
motion (FBM) \cite{mandelbrot_fbm}, L\'evy walk (LW) \cite{levy_flight,
chechkin_levyflight, shlesinger_levywalk, zaburdaev_levywalk}, scaled Brownian
motion (SBM) \cite{lim_sbm, jeon_sbm} or annealed transient time motion (ATTM)
\cite{massignan_attm}, as detailed in section \ref{sec_andi}. We evaluate the
performance of the uncertainty estimation for the regression of the anomalous
diffusion exponent (section \ref{sec_results_reg}) and the classification of
the diffusion model (section \ref{sec_results_class}). We find
that for both classification and regression the added error estimate does not
diminish performance, such that we can still achieve results on par with the
best \emph{AnDi-Challenge} competitors. The added error estimate proves to
be highly accurate even for short trajectories, an observation that merits
a detailed investigation of its behaviour. We analyse the error prediction
behaviour depending on the diffusion model, anomalous diffusion exponent,
noise and trajectory length in order to obtain insights into the learning
process of the machine. To differentiate between error predictions due to
model uncertainties and those inherent in each model, we further analyse the
predicted uncertainties for the inference of the anomalous diffusion exponent
with known ground truth diffusion model in section \ref{sec_singlemod}.
We show that the observed dependencies can be attributed to
specific properties of the underlying diffusion models.

\subsection{Regression}
\label{sec_results_reg}

\begin{figure*}
\includegraphics[width=0.98\linewidth]{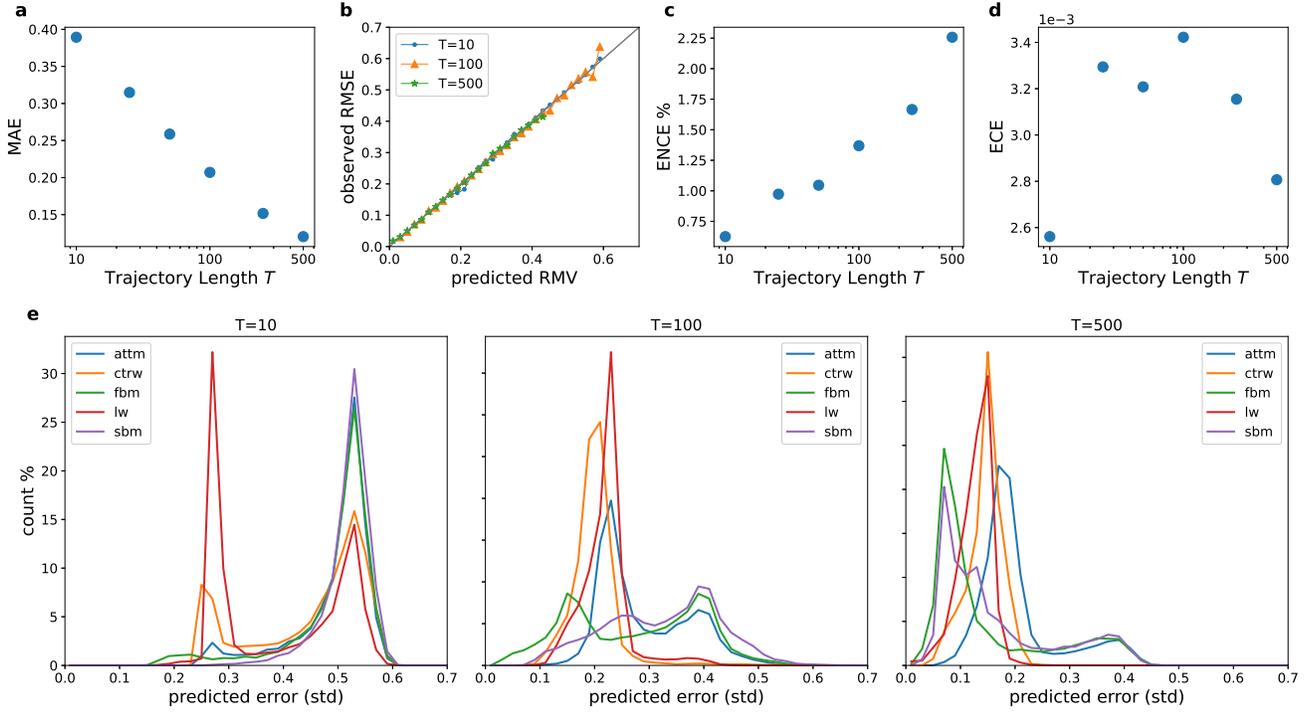}
\caption{Performance evaluation for the regression of the anomalous exponent
$\alpha$. (a) Mean Absolute Error (MAE), (b) reliability diagram
and Expected (c) Normalised and (d) non-normalised Calibration Error
(ENCE/ECE)\cite{levi_ence} achieved by \textit{Multi-SWAG} (see Supplementary
Information for detailed definitions). Results are plotted for
different trajectory lengths $T$ by averaging over $10^5$ test trajectories
each. The MAEs (a) show a decreasing trend with trajectory length, with
results close to those achieved in the \emph{AnDi-Challenge}, reaching an MAE of
0.14 for $T=500$. To judge the error prediction performance, the reliability
diagram (b) depicts the observed root mean squared error (RMSE) as a function
of the predicted root mean variance (RMV), showing that even for very short
trajectories $T=10$ an error prediction close to the ideal (grey line) is
achieved.  The reliability diagram can be summarised in a single value using
the ECE/ENCE.  The ENCE/ECE characterises the mean difference between predicted
and observed errors, either normalised to obtain a relative error (ENCE) or as
an absolute (ECE). As visible in (b), we obtain good error predictions with
an ENCE between $0.6-2.3\%$ depending on the trajectory length. The increase
in ENCE with trajectory lengths can be attributed to the decrease in MAE
(and therefore predicted errors), while the unnormalised ECE only shows a
slight trend of decreasing with trajectory length. The low ECE for $T=10$
is due to the high number of trajectories predicted with near maximal error.\\
(e) Predicted error histogram for inferring the anomalous diffusion
exponent $\alpha$ when the underlying model is unknown. The figure shows
the distribution of the error as predicted by \emph{Multi-SWAG} trained on all
models. Each subplot shows the results for a different trajectory length $T$,
as obtained from predictions on $10^{5}$ trajectories.}
\label{fig_regece}
\end{figure*}

In order to quantify the performance of our \textit{Multi-SWAG}
\cite{wilson_multiswag} models we test them on a new set of computer
generated trajectories using the \texttt{andi-datasets} package.
For the general prediction of the anomalous diffusion exponent $\alpha$ we
obtain results comparable to the best participants in the \emph{AnDi-Challenge}
\cite{munozgil_andich_end,aghion,krog_bay1,park_bay2,thapa_bay3,argun,gentili,li,verdier,manzo_elm,granik,garibo,janczura,kowalek,loch-olszewska,bo}.
The achieved mean average error for different trajectory lengths in figure
\ref{fig_regece}a shows an expected decreasing trend with trajectory length.

To analyse the performance of the error prediction we use a reliability
diagram \cite{degroot_forecast,guo_reliability, naeini_ece} in figure
\ref{fig_regece}b. {The figure depicts the observed root mean squared
error (RMSE) from the ground truth exponent as a function of the predicted
root mean variance (RMV) (see Supplementary Information for detailed
definitions)}. Grouping together predictions within a range of $0.02$, we see
results close to the ideal of coinciding predictions and observations. As
is to be expected, longer trajectories show smaller predicted errors, yet,
the higher errors for very short trajectories of only $10$ time steps are
still predicted remarkably well. The results of the reliability diagram
can be summarised using the Expected Normalised Calibration Error (ENCE)
{\cite{levi_ence}}, which calculates the normalised mean deviation between
observed and predicted uncertainty. Figure \ref{fig_regece}c shows
a low ENCE between $0.6\%$ and $2.3\%$, which increases with trajectory
length. This increase can be attributed to the decrease in predicted standard
deviations, which results in a higher normalised error due to the fact that
the unnormalised expected calibration error (ECE) only shows a slight decrease
with trajectory length, as can be seen in figure \ref{fig_regece}d.

In order to better understand how the network obtained these predictions,
it proves useful to observe the frequency of predicted standard deviations
in figure \ref{fig_regece}e. The histograms there show how often which error
is predicted for different ground truth models.

For very short trajectories ($T=10$) we observe a split of the predictions
into two peaks. This observation can be attributed to the different priors
of the ground truth models. If the network can confidently identify the
trajectory as belonging to one of the only sub-/superdiffusive models
(CTRW/LW/ATTM), it can predict (and achieve) a smaller error due to the
reduced range of possible $\alpha$-values. From the different heights of this
second peak, we can also conclude that, for very short trajectories, LW is
easier to identify than CTRW or ATTM. This is likely due to the fact that
LWs have long structures without a change in direction, that can be fairly
easily identified, while CTRWs with long resting times will be particularly
camouflaged by the noise and ATTMs without jumps in the diffusivity will be
indistinguishable from normal diffusion. Other than identifying the model the
network does not seem to gain much information from these short trajectories
as the two peaks are close to the maximum predicted errors one would expect
with respect to the priors. FBM trajectories, however, are an exception to
this, as one may already see a small amount of very low predicted errors,
which will be further studied in section \ref{sec_singlemod}.

When increasing the trajectory lengths we see lower error predictions for
all models. Both FBM and SBM achieve lower predicted errors than the other
three models, despite the larger range of $\alpha$, which may be attributed
to the fact that they do not rely on hidden waiting times, in contrast to
the other three models. While we see FBM's accuracy increasing faster than
SBM's at the beginning for $T=100$, we obtain similar predicted errors for
the two models for $T=500$. This may be caused by SBM being highly influenced
by noise (see section \ref{sec_singlemod}) and thus easier to be confused with
ATTM, since both feature a time dependent diffusivity. The errors introduced
by model confusion can also be observed in the persisting second peak. As we
will see below, this peak can be understood as a property of ATTM. An ATTM
trajectory with no jumps in diffusivity, which will occur more often for very
subdiffusive trajectories (small $\alpha$), will be indistinguishable from
normal diffusion with $\alpha=1$, thereby introducing a large error. Due
to the uncertainty in the underlying model this predicted error is also
present for both FBM and SBM, both exhibiting ordinary Brownian Motion for
$\alpha=1$.

Analogously to the other models the predicted error for LW and CTRW reduces
with increased trajectory length. CTRW shows less error than LW for $T=100$,
which may be attributed to the smaller prior used for the CTRW trajectories
$0.05\leq\alpha\leq1$ compared to LW $1<\alpha\leq2$. For $T=500$ this
difference vanishes, as the importance of different priors decreases with
better accuracy, and we even see a slightly lower predicted error for LW.

\subsubsection{Single Model Regression}
\label{sec_singlemod}

\begin{figure}
\includegraphics[width=0.98\linewidth]{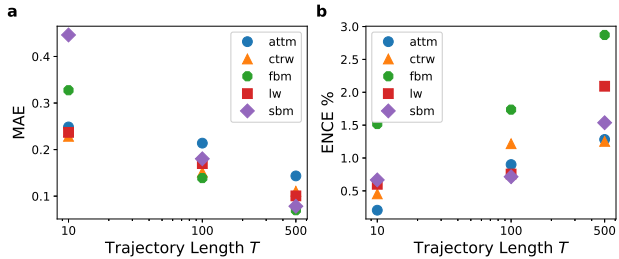}
\caption{Performance evaluation for regression when the underlying model is
known. (a) Mean Absolute Error (MAE) and (b) Expected Normalised Calibration
Error (ENCE)\cite{levi_ence} achieved by the \textit{Multi-SWAG} models
trained on only one model, plotted for different trajectory lengths $T$
by averaging over $5\times10^4$ (FBM, SBM) or $4\times10^4$ (ATTM, LW, CTRW)
test trajectories each. The ENCE characterises the mean difference between
predicted and observed errors. As was the case for the unknown ground truth
model (figure \ref{fig_regece}), we can achieve a small calibration error
below $3\%$. The MAE shows the expected results with regards to the histograms
in figure \ref{fig_histsing}.}
\label{fig_regecesing}
\end{figure}

\begin{figure*}
\includegraphics[width=\linewidth]{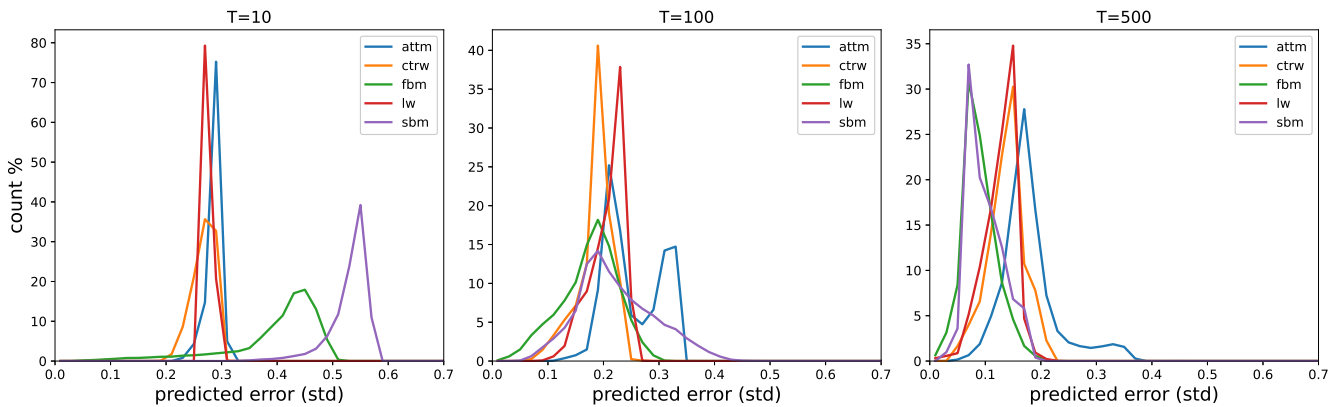}
\caption{Predicted error histogram for inferring the anomalous diffusion
exponent when the underlying model is known. The figure shows the distribution
of the error as predicted by neural networks trained individually for each
model. The histograms are obtained from predictions on $5\times 10^{4}$
(FBM/SBM) or $4\times 10^{4}$ (ATTM/LW/CTRW) trajectories.}
\label{fig_histsing}
\end{figure*}

\begin{figure*}
\includegraphics[width=0.98\linewidth]{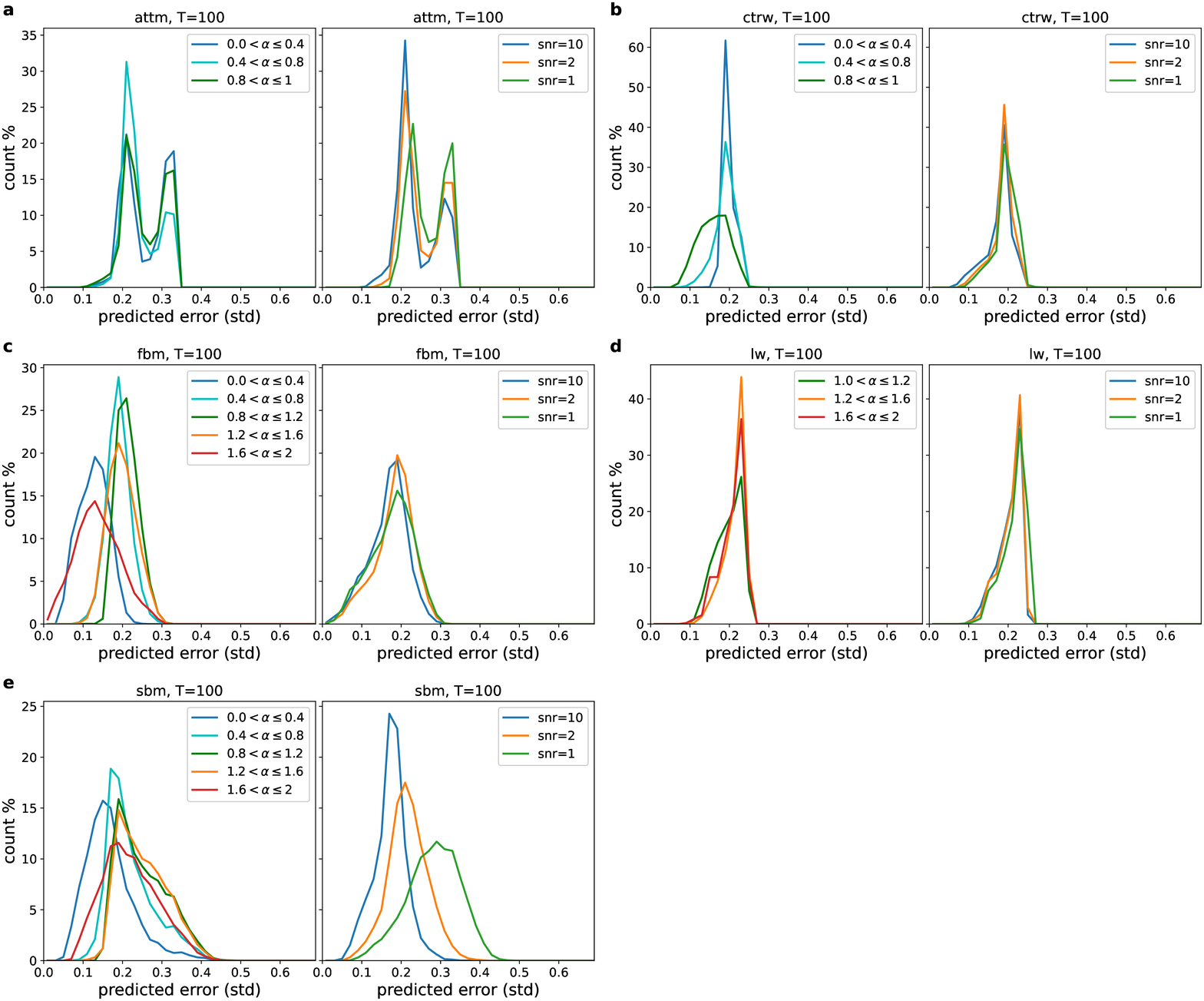}
\caption{Predicted error histogram for known ground truth model, split by
exponent and noise for trajectories of length $T=100$. Each of the figures
[(a) FBM, (b) SBM, (c) ATTM, (d) CTRW, (e) LW]
shows the results for one of the five different ground truth models,
obtained from predictions on $5\times 10^{4}$ (FBM/SBM) or $4\times 10^{4}$
(ATTM/LW/CTRW) trajectories.}
\label{histo}
\end{figure*}

In order to differentiate between errors originating from the model
uncertainty and errors specific to an individual model, it proves useful
to perform a regression of the anomalous diffusion exponent $\alpha$ on
only a single diffusion model with networks trained on only that model. As
before we are able to obtain small ENCEs below $3\%$, as seen in figure
\ref{fig_regecesing}. Due to this low calibration error the achieved MAEs
in figure \ref{fig_regecesing} largely resemble the predicted errors in the
histograms in figure \ref{fig_histsing}, which will be discussed in detail
in the following. In addition we analyse the change in predicted errors with
respect to the ground truth exponent and the noise, using the histograms in
figures \ref{histo}a to \ref{histo}e for trajectories of length $T=100$,
as well as supplementary figure S1 for lengths $T=10$ and $500$.

\paragraph{FBM.}
As one expects, due to the larger prior, FBM's error predictions for very short
trajectories ($T=10$) are larger than the three exclusively sub- or superdiffusive
models. Compared to SBM and the performances for unknown ground truth models in
figure \ref{fig_regece}e, these errors are, however, remarkably low, {showing
that, while the correlations for very short trajectories were not noticeable
enough to identify them as FBM above, they are enough to significantly improve
the performance when they are known to be FBM trajectories.} Additionally
one may also notice a small percentage of trajectories assigned with very low
predicted error, which can also be seen for longer trajectories but is less
noticeable. As before, we see that the predictions quickly improve for longer
trajectories and ultimately reach better results than for ATTM, LW, or CTRW.

By studying the dependence of the predicted error on the ground truth exponent
in figures \ref{histo}a and S1, we can attribute
the low error predictions to the very super-/subdiffusive trajectories,
for which correlations are apparent. This feature occurs despite of the fact
that for short trajectories only the superdiffusive trajectories contribute,
which is likely caused due to anticorrelations in short trajectories being
similar to noisy trajectories. Concerning the dependence on noise we only
see a slight increase in the predicted accuracy for lower noise regardless
of the trajectory length, although the possibility of high noise likely
influences the predictions, as explained above.

\paragraph{SBM.}
Similar to FBM, due to the large prior, SBM trajectories start with high
error predictions for very short trajectories in figure \ref{fig_histsing}. In
contrast to FBM, however, these predictions are much higher, since a change
in diffusivity will be hard to detect for few time steps. When increasing
the lengths, the predictions improve, getting close to those for FBM for
$T=500$. Similar to above, we also observe a noticeably broad distribution
of errors, this time however to the right side of the peak. We can explain
this broadness by examining the noise dependence of the predictions in
figure \ref{histo}b (and S1). We see a
large difference between predicted errors depending on noise. For example,
for length $T=100$ we obtain a mean predicted standard deviation of $\approx
0.032$ for low noise ($\text{snr}=10$) and $\approx 0.082$ for high noise
($\text{snr}=1$), more than doubling the error. We can attribute this effect
due to the influence of static noise on a trajectory, whose increments
increase/decrease over time for super-/subdiffusive trajectories. This will
effectively hide part of the data under high noise, reducing the number
of effectively useful data points.

When observing the dependence of the
predicted error on the ground truth exponent in figure \ref{histo}b we
can see better predictions for the more pronouncedly sub- and superdiffusive
cases for length $T=100$, showing that despite the fact that part of
these trajectories are hidden under the noise, the large increase/decrease in
diffusivity still makes these trajectories easier to identify. One should also
keep in mind that while these will be very noisy at one end, they will also
be less noisy at the other end. The network does, however, assign a lower
predicted error for subdiffusive trajectories than for superdiffusive ones,
for which the difference increases for larger snr. This may indicate, that
the subdiffusive decrease in diffusivity ($\propto 1/t^{1-\alpha}\to 1/t$
for $\alpha\to 0$) is easier to identify than the superdiffusive increase
($\propto t^{\alpha-1}\to t$ for $\alpha\to 2$). The former will have a larger
portion of the trajectory hidden under the noise with a steep visible decrease
at the beginning, while the latter will increase more slowly, leading to
a smaller hidden portion but also making the non-hidden part less distinct
and the transition between more ambiguous.

\paragraph{ATTM.}
In figure \ref{fig_histsing} we see a behaviour for ATTM similar to what was
discussed in the previous section. This time the histogram starts for short
trajectories as a single peak close to the maximum prediction possible with
respect to the prior. With increasing length the peak splits into two peaks,
where the second peak, as discussed above, originates from subdiffusive
ATTM trajectories with few or no jumps in the diffusivity. This second
peak decreases in volume for very long trajectories, since observing no
jumps becomes rarer and it becomes easier to identify the still occurring,
albeit small, jumps in normal-diffusive ($\alpha=1$) ATTM trajectories. The
second point should also be the reason why the right peak is less pronounced
than in the case of unknown underlying model in figure \ref{fig_regece}e,
as it is easier to confuse subdiffusive ATTM with normal-diffusive FBM/SBM
than with normal-diffusive ATTM.

For the $\alpha$-dependence in figures \ref{histo}c and
S1 we can see that, as expected, the right
peak is more pronounced for sub- and normal-diffusive trajectories. For
length $T=500$ (figure S1) we also see that the
lowest errors originate from close to normal-diffusive trajectories, as
these will exhibit more jumps and thereby allow to identify more waiting
times. As for the influence of the noise, in figure \ref{histo}c
(S1) we see a slight increase of the uncertainty with
higher noise, as well as the right peak being more pronounced for higher
noise, likely due to the fact that the noise obscures the smaller jumps
occurring in normal-diffusive ATTM.

\paragraph{CTRW.}
As seen in figure \ref{fig_histsing} CTRW shows a single peak, whose location
shifts to lower predicted errors with increasing trajectory length. When
examining the dependence on the ground truth $\alpha$ value and noise in
figures \ref{histo}d and S1, one can see that
an increase in the noise will have little effect on the predictions, only
leading to a slight increase in the predicted error. The largest difference is
observed for very short trajectories in figure S1,
likely for the fact that the low noise here allows one to detect the
very few jumps in the short trajectories. The exponent $\alpha$, however,
has a higher influence on the error predictions. One can observe that the
predicted error will be smaller for exponents closer to normal diffusion,
arguably as more jumps occur in this case.

\paragraph{LW.}
The LW evaluation in figure \ref{fig_histsing} exhibits similar behaviour to
the CTRW, showing a single peak shifting toward lower predicted errors. As
discussed above the predictions for LW are slightly worse than for CTRW in
the beginning, which we attribute to the difference in the prior.  In figures
\ref{histo}e and S1, we see little to no
influence of the noise on the error predictions. From these figures one may
also obtain a similar, though much less pronounced, behaviour in dependence
of the ground truth $\alpha$ as for CTRW. As was the case there we see lower
predictions for exponents close to normal diffusion, as more hidden waiting
times can be observed. Interestingly in figure S1
we see that for long trajectories the predicted error will also be reduced
for very superdiffusive trajectories. In part, this can be attributed to
the distinct ballistic $\alpha=2$ LW, but should also be caused by the noise
as superdiffusive LW with a few very long jumps is, in contrast to CTRW with
few jumps, not highly influenced by noise.

\subsection{Classification}
\label{sec_results_class}

\begin{figure*}
\includegraphics[width=0.98\linewidth]{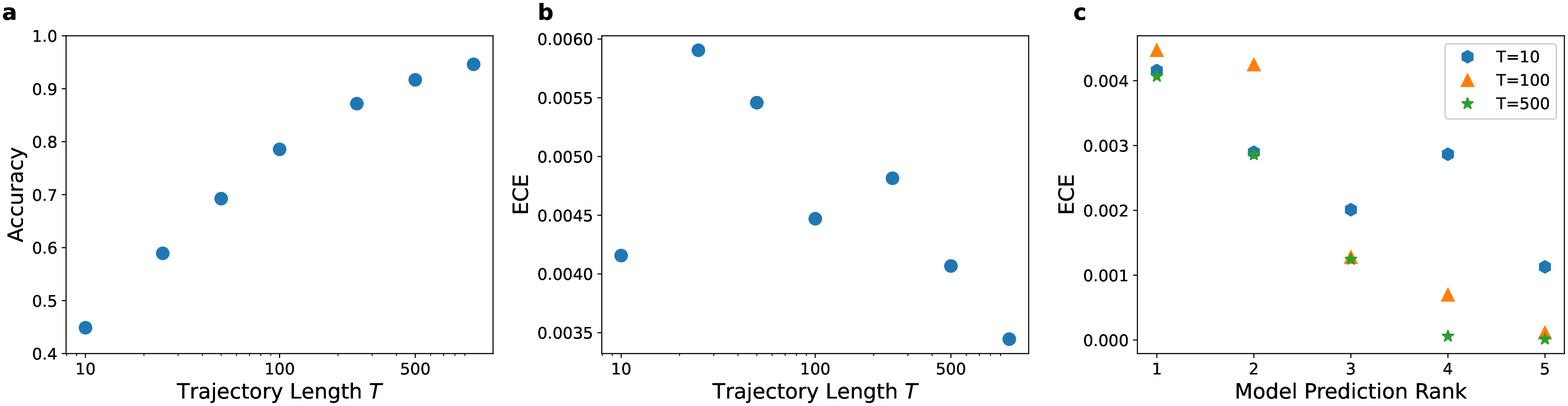}
\caption{Performance evaluation for the classification of the diffusion model.
(a) Total accuracy and (b) Expected Calibration Error
(ECE) \cite{guo_reliability, naeini_ece} (see Supplementary Information
for detailed definitions) achieved by \textit{Multi-SWAG}, plotted for
different trajectory lengths \textit{T} by averaging over $10^5$ trajectories
each. The ECE describes the difference one may expect between the predicted
confidence and the observed accuracy. As before we achieve a low calibration
error between 0.3\% and 0.6\%. The classification accuracy improves the
longer the trajectory, achieving results similar to the best scoring models
in the \emph{AnDi-Challenge} \cite{munozgil_andich_end}.\\
(c) Expected Calibration Error (ECE) \cite{guo_reliability, naeini_ece}
achieved for lower-ranked predictions, meaning those models that were not
assigned the highest confidence. A prediction of rank $i$ corresponds to
the output with the $i$th highest confidence. Even these predictions show
low calibration errors below 0.5\%. The vanishing ECE for the 4th and lower
ranked predictions of long trajectories are caused by them being correctly
assigned a 0\% probability.}
\label{fig_class_ece}
\end{figure*}

\begin{figure*}
\includegraphics[width=0.98\linewidth]{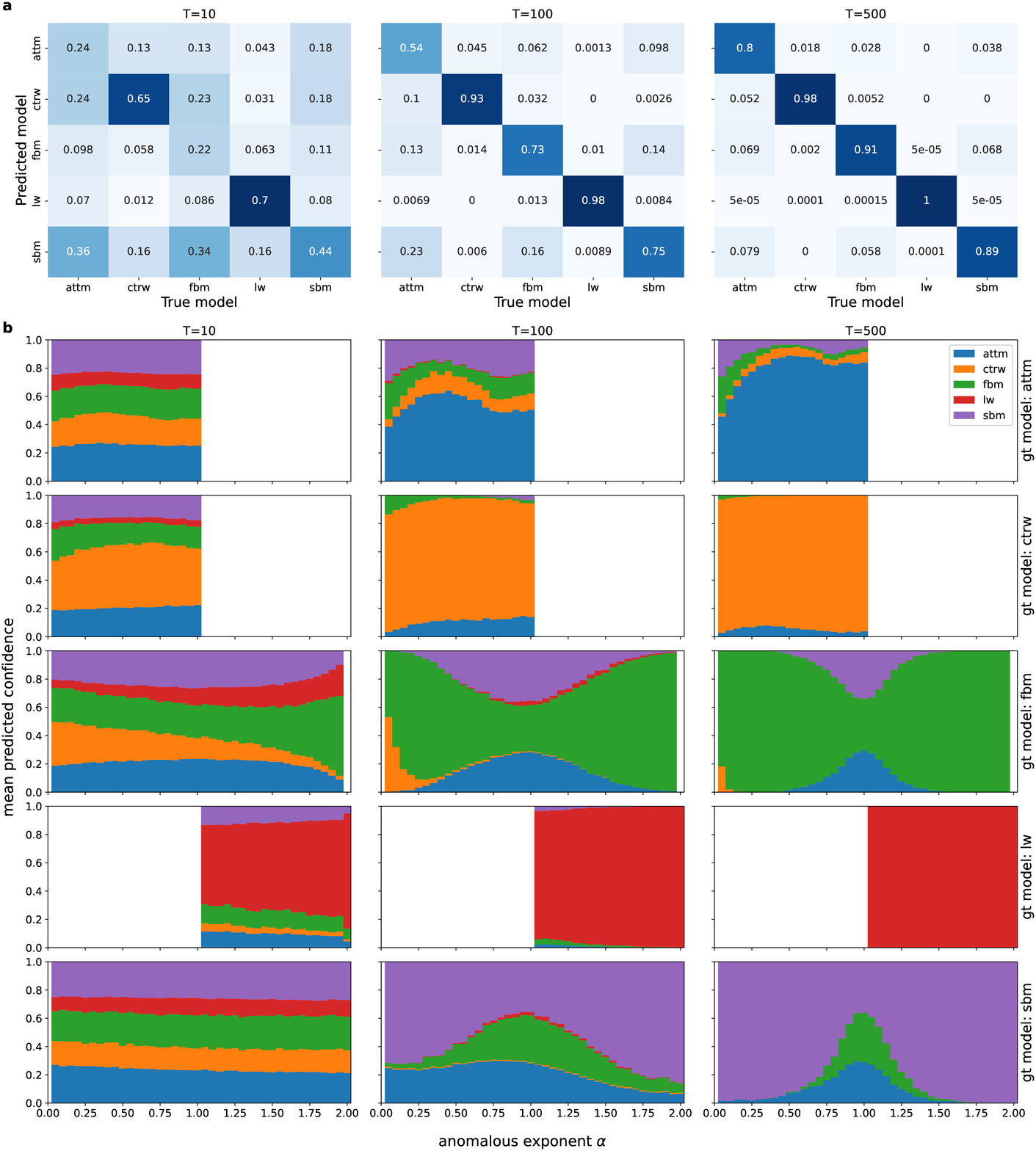}
\caption{Analysis of the classification behaviour.
(a) Confusion matrices for different trajectory lengths. The entries
indicate the relative frequency of model predictions (row) given a ground
truth model (column). The matrices are obtained from \emph{Multi-SWAG}
predictions on a total of $10^5$ test trajectories for each length.\\
(b) Mean confidences for different ground truth (gt) models in
dependence of the ground truth anomalous diffusion exponent $\alpha$. From the
figure one can infer the confidence for the different models shown as coloured
bars as assigned by \textit{Multi-SWAG}. The illustrations are plotted for
different trajectory lengths by averaging over a total of $10^5$ trajectories
for each length, translating to $2\times 10^4$ trajectories for each ground
truth model and length.}
\label{fig_class_meanconf}
\end{figure*}

Complementing the discussion of the regression in section
\ref{sec_results_reg}, we now evaluate the trained
\textit{Multi-SWAG} models on the test data set. The
achieved accuracies depicted in figure \ref{fig_class_ece}a are
in line with the best performing participants of the \emph{AnDi-Challenge}
\cite{munozgil_andich_end,aghion,krog_bay1,park_bay2,thapa_bay3,argun,gentili,
li,verdier,manzo_elm,granik,garibo,janczura,kowalek,loch-olszewska,bo}.
As one would expect the achieved accuracy increases with trajectory length,
starting from $44.9\%$ for $T=10$ and reaching $91.7\%$ for $T=500$. In
figure \ref{fig_class_ece}b we also see a very good performance for error
prediction, the expected calibration error only ranging from 0.3 to 0.6
percentage points. The ECE generally shows a decreasing trend with increasing
trajectory length, although very short trajectories of $T=10$ also achieved a
low ECE, likely due to a high number of trajectories predicted with very low
confidences.  Remarkably even the confidences of the lower-ranked predictions,
relating to those models that were not assigned the highest confidence,
achieve similarly low ECEs in figure \ref{fig_class_ece}c.

To further analyse the performance and error prediction, we show
the confusion matrices in figure \ref{fig_class_meanconf}a and the
mean predicted confidences in figure \ref{fig_class_meanconf}b. The
confusion matrices depict how often a model is predicted given a
specific ground truth models, thereby showing how often and with
which model each model is confused. As such matrices do not consider the
predicted confidences and have already been thoroughly examined in other works
\cite{munozgil_andich_end,aghion,krog_bay1,park_bay2,thapa_bay3,argun,gentili,li,verdier,manzo_elm,granik,garibo,janczura,kowalek,loch-olszewska,bo},
we will focus our investigation on the second figure \ref{fig_class_meanconf},
which illustrates the mean predicted confidences of each model for different
ground truth models in dependence of the true anomalous diffusion exponent
$\alpha$. Note that while the mean confidence will in part reflect the
predictions in the confusion matrix, this quantity also provides additional,
complemtary information, as
the confusion matrix only considers the models with the highest membership
score. In the following we analyse the results for different ground truth
models.

\paragraph{ATTM.}
ATTM trajectories generally show the worst classification performance {of the
range of models studied here}. For very short trajectories ($T=10$) we see that
the mean confidence splits among all models with the lowest probabilities being
assigned to the {exclusively} superdiffusive LW. Reflecting the confusion
matrix, the confidences for SBM are the highest, likely due to both SBM
and ATTM featuring a time dependent diffusivity. For longer trajectories
we see the confidences for FBM and SBM rise for lower $\alpha$, which, as
explained above, can be attributed to that fact that ATTM without jumps is
indiscernible from ordinary Brownian motion. The confusions for CTRW, which
are most present for moderately subdiffusive to normal-diffusive trajectories,
can be attributed to the fact that both models feature hidden waiting times
and short periods of high diffusivity in ATTM appear, similar to jumps in CTRW.

\paragraph{CTRW.}
Reflecting the high accuracies in the confusion matrices, we observe
high confidences for CTRW for longer trajectories ($T\geq 100$). For very
subdiffusive trajectories we see an increase in the predicted probability
for FBM, which can be explained by the fact that CTRWs without jumps solely
consist of noise, which corresponds to an FBM trajectory with $\alpha=0$. We
can also observe a similar confusion behaviour between ATTM and CTRW as was
described for ATTM. For very short trajectories ($T=10$) the confidences
for CTRW are relatively high as compared to the other ground truth models, and
they increase with higher anomalous diffusion exponent, which we attribute to
the increase in jump frequency with higher $\alpha$. Here confidences for
models other than CTRW are split between ATTM, FBM, and SBM with only small
confidences assigned to the solely superdiffusive LW.

\paragraph{FBM.}
Similarly to what we described in section \ref{sec_results_reg}, for shorter
trajectories, we see a large difference in FBM confidences for very sub- and
superdiffusive $\alpha$. We there hypothesised this difference to be caused by
the inability to discern very subdiffusive trajectories from noise. This can
be confirmed here, as subdiffusive trajectories show the highest confusion
with CTRW, which without jumps solely consists of noise. For very short
trajectories we see an increase in LW confidence with increasing $\alpha$,
likely due to highly correlated, very short FBM trajectories looking similar
to LW trajectories without jumps. For longer trajectories one can observe low
FBM confidence at and around $\alpha=1$, which is caused by FBM's convergence
to normal diffusion and leads to split uncertainties between FBM, SBM, and
ATTM. One should note that the ATTM confidences here would not correspond to
a normal-diffusive ATTM but rather to a strongly subdiffusive ATTM without
jumps in diffusivity, as is evidenced by the mean confidences for ATTM ground
truth trajectories.

\paragraph{LW.}
In accordance to the high accuracies observed in the confusion matrices, the
mean confidences for LW are high even for relatively short trajectories. These
high confidences occur, as LW is easily identifiable even with few jumps. In
fact the increase in confidence with rising anomalous diffusion exponent
suggests that LW trajectories are easier to identify when fewer jumps occur,
{which is in contrast to ATTM/CTRW, which both feature a decrease in model
confidence with fewer jumps}. One should also note the jump in confidence
caused by ballistic LW ($\alpha=2$).

\paragraph{SBM.}
As was the case for FBM, for longer SBM trajectories we see the same
confusion pattern between SBM, ATTM, and FBM at and around normal diffusion
$\alpha=1$. However, we also see relatively high assigned confidences for
ATTM for subdiffusive trajectories, which we again attribute to both models
featuring time dependent diffusivities. We see low confidences for SBM for
very short trajectories, likely due to a change in diffusivity not being
noticeable for so few data points.

In the supplementary figures S2a-c we include error
histograms similar to those used for regression. These resemble the already
discussed behaviour and indicate in addition that the distribution of
predicted errors often features a large number of trajectories predicted
with high confidences of 95\% to 100\%.

\section{Discussion}
\label{sec_last}

The \emph{AnDi-Challenge\/} demonstrated the power of a rich arsenal of
successful machine learning approaches to analyse anomalous diffusion
trajectories. These proposed models, however, all suffered from a lack of
explainability due to the Black Box problem, {providing answers without
explanation}, which also leads to an uncertainty in the reliability and
usefulness of the approaches for real-world systems.

Here we expanded the successful machine learning solutions featuring in the
\emph{AnDi-Challenge\/} by adding a reliability estimate to the predictions of
the machine. This estimate was obtained by modelling aleatoric and epistemic
uncertainties in the model, the latter by using a Bayesian machine learning
techniques called \textit{Multi Stochastic Weight Averaging Gaussian}. We
showed that the resulting model is able to provide accurate error estimates
even for very uncertain predictions when tested on separate, but identically
distributed, test data sets. It was also demonstrated that these uncertainty
predictions provide an additional tool to understand how machine learning
results are obtained. By analysing the prediction behaviour with respect
to diffusion model, noise, anomalous diffusion exponent, and trajectory
length, we were able to relate its cause to the properties of the underlying
anomalous diffusion models. This analysis also indicated that a network
trained to predict the anomalous diffusion exponent will already learn to
differentiate between the anomalous diffusion models. In our study we also
introduced the mean confidence diagrams and showed that they provide vital
information complementary to confusion matrices.

For future works testing the \textit{Multi-SWAG} models on diffusion data
whose dynamics are not included in the training set will be an interesting
field of study. Such data may include trajectories generated with different
diffusion models, a subordination or superposition of models or with changing
models. Results here will indicate, what behaviour one should expect when using
these models on experimental data, as such data will rarely exactly follow the
theoretical models. Naturally though this can and should not replace the need
to test the developed methods here as well. Similarly it might
be of interest to analyse the results obtained when applying these methods to
``poisoned'' (faulty) test data, e.g., when non-Gaussian errors contaminate
the data, non-trained stochastic mechanisms are included, or the analysed
time series have missing points. As one would expect, this leads to a higher
predicted error due to the epistemic uncertainty, as described in section
\ref{sec_bdl}. Quantifying such errors systematically will be an interesting
question for the future. We also mention that applying the used BDL methods
to the feature-based approaches for decoding anomalous diffusion data brought
forth recently {\cite{janczura,kowalek,loch-olszewska,pinholt_fingerprint}}
and analysing error prediction performance as well as the impact of the
different features on these error predictions, could also provide interesting
insights. Another interesting avenue could be provided by the third task of
the \emph{AnDi-Challenge}, which consisted of predicting the change point of a
diffusion trajectory switching models and/or exponent. Recent studies suggest
that sequence to sequence networks, predicting trajectory properties at each
time step, are suited to solve this task \cite{munozgil_andich_end}. Here
BDL might provide an advantage in addition to the error estimate, as one
would expect the predicted uncertainty to maximise at the change point and
thereby simplify its determination.

\section{Methods}
\label{sec_mainmethods}
\subsection{Anomalous Diffusion Models}
\label{sec_andi}

For comparability the models considered in this work are the same
as those in the \emph{AnDi-Challenge} \cite{munozgil_andich_start,
munozgil_andich_end}. The trajectories are generated from one of the 5 models
below, all producing an MSD of the form $\langle \mathbf{r}^2(t)\rangle\propto
K_\alpha t^\alpha$. Examples for each model are shown in figure
\ref{fig_trajplot}.

\paragraph*{CTRW.}
The continuous-time random walk (CTRW) is defined as a random walk, in
which the times between jumps and the spatial displacements are stochastic
variables \cite{montroll_ctrw,hughes_ctrw,weissmann_ctrw}. In our case, we
are considering a CTRW for which the waiting time distribution $\Psi(\tau)$
features a power law tail $\Psi(\tau)\propto\tau^{-1-\alpha}$ with scaling
exponent $0<\alpha<1$, thereby leading to a diverging mean waiting time
$\int_0^\infty \tau\Psi(\tau) d\tau = \infty$. The spatial displacements
follow a Gaussian law.

\paragraph*{LW.}
The L{\'e}vy walk (LW) is a special case of a CTRW. As above we consider
power law distributed waiting times $\Psi(\tau)\propto\tau^{-1-\sigma}$,
but the displacements are correlated, such that the walker always moves with
constant speed $v$ in one direction for one waiting time, randomly choosing
a new direction after each waiting time. One can show that this leads to an
anomalous diffusion exponent $\alpha$ given by \cite{zaburdaev_levywalk}
\begin{equation}
\alpha=\left\{\begin{array}{ll}2&\mbox{ if } 0<\sigma<1 \mbox{ (ballistic
diffusion)}\\3-\sigma & \mbox{ if } 1<\sigma<2 \mbox{ (superdiffusion)}.
\end{array}\right. 
\end{equation}

\paragraph*{FBM.}
Fractional Brownian motion (FBM) is characterised by a long-range correlation
between the increments. It is created by using fractional Gaussian noise
for the increments given by
\begin{equation}
\langle\xi_{fGn}(t)\xi_{fGn}(t+\tau)\rangle\sim\alpha(\alpha-1) K_\alpha \tau^{\alpha-2}
\end{equation}
for sufficiently large $\tau$, where $\alpha$ is the anomalous diffusion exponent and $K_\alpha$ is the generalised diffusion constant \cite{mandelbrot_fbm}.

\paragraph*{SBM.}
Scaled Brownian motion (SBM) features the time dependent diffusivity $K(t)=
\alpha K_\alpha t^{\alpha-1}$, equivalent to the Langevin equation
\begin{equation}
\frac{dx(t)}{dt}=\sqrt{2K(t)}\xi(t),
\end{equation}
where $\xi(t)$ is white, zero-mean Gaussian noise \cite{jeon_sbm}.

\paragraph*{ATTM.}
Similar to SBM, the annealed transient time motion (ATTM) features a diffusion
coefficient $D$ varying over time. But in contrast to SBM, the change in
diffusivity is random in magnitude and occurs instantaneously in a manner
similar to the jumps in a CTRW. Here we consider diffusion coefficients
sampled from the distribution $P(D)\propto D^{\sigma-1}$ and use a delta
distribution of waiting times $P(\tau)\propto \delta(\tau-D^{-\gamma})$,
with $\sigma<\gamma<\sigma+1$. As shown in \cite{massignan_attm}, this leads
to subdiffusion with $\alpha = \sigma/\gamma$.

We use the \texttt{andi-datasets} Python package for the implementation of
these models \cite{andidataset_python}. In an effort to simulate conditions
closer to experimental situations, all data are corrupted by white Gaussian
noise with the signal to noise strength ratio $\text{snr}\in\{1,2,10\}$. Given
the trajectory $x_t$, we obtain the noisy trajectory $\tilde{x}(t)=x(t)+\xi(t)$
with the superimposed noise
\begin{equation}
\xi(t) \sim \frac{\sigma_{\Delta x}}{\text{snr}} \mathcal{N}(0,1)\text{,}\label{eq_noise}
\end{equation}
where $\sigma_{\Delta x}$ is the standard deviation of the increment process
$\Delta x (t) = x(t+1)-x(t)$. We consider trajectories generated with anomalous
diffusion exponents $\alpha\in\{0.05,0.10,...,1.95,2\}$. Note however that only
SBM is applied to the whole range of $\alpha$ values. CTRW and ATTM are only
sub- or normal-diffusive ($\alpha\leq1$), LW is superdiffusive ($\alpha> 1$)
and ballistic ($\alpha=2$) FBM is not considered here. This entails that data
sets with a mixture of models cannot be equally distributed with respect to
the anomalous diffusion exponents and underlying models at the same time.
In this work we choose the prior distributions of models and
exponents such that they conform with those used in the \emph{AnDi-Challenge},
where the priors were chosen to simulate no prior-knowledge for the given task.
This entails that the data set used for the classification tasks is equally
distributed with respect to models but not among anomalous diffusion exponents,
and vice versa for the data set used for the regression of $\alpha$.
Subdiffusive trajectories are therefore overrepresented in the classification
data sets, while FBM and SBM will be overrepresented for regression.

\subsection{Uncertainties in Deep Learning}
\label{sec_bdl}

In short, a neural network in deep learning is a function approximator,
where the output $f_\theta(x_i)$ of the neural network given inputs $x_i$
is optimised to minimise some loss function $\mathcal{L}$. This is achieved
by fitting the function parameters (weights) $\theta$ of the neural network,
usually by utilising the stochastic gradient descent algorithm or a variation
of it {\cite{bottou_sgd}}.

In Bayesian Deep Learning, one differentiates between two major types of
uncertainty named aleatoric and epistemic uncertainty \cite{kiureghian_aleaepi,
kendall_whatuncert}.

\subsubsection*{Aleatoric Uncertainty}

Aleatoric uncertainty refers to the uncertainty inherent in the system
underlying the data, caused, for example, by noise or an inherent stochasticity
of the system. This kind of uncertainty needs to be included in the output
of the neural network model. We then minimise the negative log likelihood loss
\begin{equation}
\mathcal{L}_{\text{nll}}=-\sum_i\log p(\hat{y}_i|f_\theta(x_i)),
\end{equation}
where $\hat{y}_i$ is the target output and $f_\theta(x_i)$ is the prediction
of the neural network given input $x_i$ and weights $\theta$ \cite{nll_loss}.

For regression problems, the commonly used models output only a predicted
value and optimise the network to minimise either the mean absolute error or
the mean squared error \cite{wang_lossfunctions}. In order to model aleatoric
uncertainty we modify the network to output mean and variance of a Gaussian
predictive distribution, instead of just predicting a single value
(while a Gaussian distribution will often not be a precise approximation, it
suffices to obtain well calibrated estimates for the standard deviation).
When $p(\hat{y}_i|f_\theta(x_i))\sim \mathcal{N}_{\mu_i,\sigma_i}(\hat{y}_i)$,
we minimise the negative log likelihood, which becomes the Gaussian negative
log likelihood loss
\begin{equation}
\mathcal{L}_{\mathrm{gnll}}=\sum_i\frac{1}{2}\left(\log(\sigma_i^2)+\frac{||
\mu_i-\hat{y}_i||^2}{\sigma_i^2}\right)+\mathrm{const}, 
\end{equation}
where $\mu_i$ and $\sigma_i$ are the mean and variance outputs of the neural
network for input $x_i$ \cite{gnll_loss}.

The commonly used models for classification already output an aleatoric
error. We train the model to {output membership scores for each class in a
so called \textit{logit} vector} $z_i=f_\theta(x_i)$, from which the class
probabilities can be obtained via a normalised exponential (\textit{softmax})
function
\begin{equation}
p_{i,k}=\frac{\exp{z_{i,k}}}{\sum_k\exp{z_{i,k}}},
\end{equation}
where $p_{i,k}$ is the predicted probability of class $k$ given input $x_i$.
From the negative log likelihood loss we then obtain the \textit{cross entropy
loss}
\begin{equation}
\mathcal{L}_{cel}=-\sum_{i,k}\hat{y}_{i,k}\log(p_{i,k}),
\end{equation}
where $\hat{y}_{i,k}$ is a binary indicator $\hat{y}_{i,k}=\delta_{j_ik}$
of the true class $j_i$ of input $x_i$.

\subsubsection*{Epistemic uncertainty and stochastic weight averaging
Gaussian (\textbf{SWAG})}

\begin{algorithm}[b]
\begin{algorithmic}[0]
\State{$\theta_0$ pre-trained weights; $\eta$ learning rate; $T$ number of training steps; $c$ moment update frequency; $K$ maximum number of columns in deviation matrix $\hat{D}$; $M$ number of Monte Carlo samples in Bayesian model averaging}
\algrenewcommand\algorithmicprocedure{\textbf{Train}}
\Procedure{SWAG}{}%{$bel(x_{t-1}),u_t,z_t$}
\State{$\bar{\theta}\gets\theta_0, \overline{\theta^2}\gets\theta_0^2$} \Comment{initialise moments}
\For{$i \gets 1$ \textbf{to} $T$}
\State{$\theta_i\gets\theta_{i-1}-\eta\nabla_\theta \mathcal{L}(\theta_{i-1})$} \Comment{SGD update}
\If{$mod(i,c)=0$}
\State{$n\gets i/c$}
\State{$\bar{\theta}\gets\frac{n\bar{\theta}+\theta_i}{n+1}$, $\overline{\theta^2}\gets\frac{n\overline{\theta^2}+\theta^2_i}{n+1}$} \Comment{update moments}
\If{number of columns$(\hat{D})=K$}
\State{remove first column in $\hat{D}$}
\EndIf
\State{append column $(\theta_i-\bar{\theta})$ to $\hat{D}$} \Comment{deviation matrix}
\EndIf
\EndFor
\State \textbf{return} $\theta_{\text{SWA}}=\bar{\theta}, \Sigma_{\text{diag}} = \overline{\theta^2}-\bar{\theta}^2, \hat{D}$
\EndProcedure
\\
\algrenewcommand\algorithmicprocedure{\textbf{Test}}
\Procedure{Bayesian Model Averaging}{}
\For{$i\gets 1$ \textbf{to} $M$}
\State{draw $\tilde{\theta}_i\sim \mathcal{N}(\theta_{\text{SWA}},\frac{1}{2}\Sigma_{\text{diag}}+\frac{\hat{D}\hat{D}^T}{2(K-1)})$}
\State{$p({y}|\text{Data}) \mathrel{{+}{=}} \frac{1}{M}p({y}|\tilde{\theta}_i)$}
\EndFor
\State \textbf{return} $p({y}|\text{Data})$
\EndProcedure
\end{algorithmic}
\caption{SWAG \cite{maddox_swag}}
\label{alg_swag}
\end{algorithm}

Epistemic uncertainty refers to the uncertainty caused by an imperfect
model, for example due to a difference between training and test data or
insufficient training data. In Bayesian Deep Learning we model this error
by assigning an uncertainty to the inferred neural network weights. If
$p(\theta|\mathcal{D})$ is the probability distribution over the weights
$\theta$ given data $\mathcal{D}$, we obtain
\begin{equation}
p({y}|x_i,\mathcal{D})=\int d\theta p({y}|x_i,\theta) p(\theta|\mathcal{D}).
\end{equation}
In practice this integral is approximated by \textit{Monte Carlo (MC)
integration} \cite{mc_sampling}
\begin{equation}
p({y}|x_i,\mathcal{D})\approx\frac{1}{M}\sum_{m=1}^{M}p({y}|x_i,\theta_m), \label{eq_mcsample}
\end{equation}
where the weights $\theta_m$ are sampled from the posterior $p(\theta|\mathcal{
D})$ and $M$ is the number of MC-samples. Mathematically this posterior is given
by \textit{Bayes' rule} \cite{kolmogorov_probtheory}
\begin{equation}
p(\theta|\mathcal{D})=\frac{p(\mathcal{D}|\theta)p(\theta)}{p(\mathcal{D})}.
\end{equation}
However as calculating the posterior becomes intractable for large
networks and data sets, we need to approximate it. For this purpose
Maddox et al. proposed a method named \textit{Stochastic Weight Averaging
Gaussian} (SWAG) \cite{maddox_swag}, which we will use in a combination
with Deep Ensembles \cite{lakshminarayanan_deep_ensemble} leading to
\emph{Multi-SWAG} as proposed by Wilson et al \cite{wilson_multiswag}. In SWAG
one interprets the stochastic gradient descent (SGD) algorithm, used
to optimise the neural network given a loss function, as approximate
Bayesian inference. SWAG estimates the first and second moment of the
running SGD iterates to construct a Gaussian distribution over the weights
$p(\theta|\mathcal{D})\sim\mathcal{N}_{\bar{\theta},\Sigma}(\theta)$.
Maddox et al. show that this Gaussian approximation suffices to capture
the local shape of the loss space around the obtained minimum.
When training a pre-trained neural network for $T$ SWAG updates, the mean
value and sample covariance are given as \cite{maddox_swag}
\begin{eqnarray}
\bar{\theta}&=&\frac{1}{T}\sum_{i=1}^T \theta_i\\
\Sigma&=&\frac{1}{T-1}\sum_{i=1}^T(\theta_i-\bar{\theta})(\theta_i-\bar{
\theta})^T.
\end{eqnarray}
As computing the full covariance matrix is often intractable, SWAG approximates
by splitting it into a diagonal covariance $\Sigma_{\text{diag}}$,
only containing the diagonal variances, and low-rank covariance
$\Sigma_{\text{low-rank}}$, which approximates the full matrix by only using
the last few update steps. The diagonal covariance is given as
\begin{equation}
\Sigma_{\text{diag}}=\text{diag}(\overline{\theta^2}-\bar{\theta}^2),
\end{equation}
where $\overline{\theta^2} = \frac{1}{T}\sum_{i=1}^T \theta_i^2$ and
the squares in $\theta_i^2,\bar{\theta}^2$ are applied element-wise. For
the low-rank covariance we first approximate $\Sigma$ using the running
estimate $\bar{\theta}_i$ after $i$ steps: $\Sigma \approx \frac{1}{T-1}
\sum_{i=1}^T (\theta_i-\bar{\theta}_i)(\theta_i-\bar{\theta}_i)^T =
\frac{DD^T}{T-1}$, where $D$ is the deviation matrix consisting of columns
$D_i=(\theta_i-\bar{\theta}_i)$. Further we only use the last $K$ columns of
$D$ in order to calculate the low rank covariance matrix. Defining $\hat{D}$
as the matrix comprised of columns $T-K+1,\ldots,T$ of $D$, we obtain
\begin{equation}
\Sigma_{\text{low-rank}}=\frac{\hat{D}\hat{D}^T}{K-1}.
\end{equation}
Thus one only needs to keep track of $\bar{\theta},\overline{\theta^2}$
and $\hat{D}$ and can sample the weights used in equation (\ref{eq_mcsample})
from the Gaussian $\mathcal{N}(\bar{\theta},\frac{1}{2}(\Sigma_\text{diag}
+\Sigma_\text{low-rank}))$. The full SWAG procedure is shown in algorithm \ref{alg_swag}.

In \emph{Multi-SWAG} one combines this SWAG algorithm with deep ensembles by training
multiple SWAG models and taking an equal amount of samples from each
\cite{wilson_multiswag}.

\subsection{\label{sec_archi}Neural Network Architecture and Training}

\begin{figure}
\includegraphics[width=0.9\linewidth]{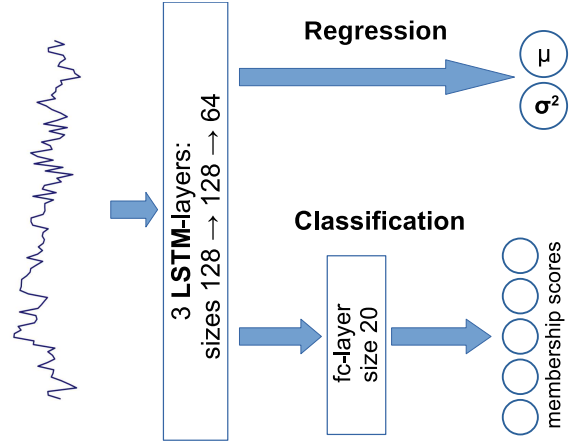}
\caption{Architecture of the used Neural Network. For both regression and
classification the network first consists of three stacked long short-term
memory (LSTM) layers \cite{hochreiter_lstm} of sizes 128, 128 and 64. For
regression the last LSTM is directly fully connected into the output layer
returning a mean $\mu$ and variance $\sigma$, while for classification the
output layer is preceded by another fully connected layer of size 20. The
architecture is inspired by the successful applications of recurrent neural
networks during the \emph{AnDi-Challenge} \cite{munozgil_andich_end,argun,
garibo,li}.}
\label{fig_architecture}
\end{figure}

Inspired by its success in the \emph{AnDi-Challenge} \cite{munozgil_andich_end}
we chose a recurrent (LSTM \cite{hochreiter_lstm}) neural network as
depicted in figure \ref{fig_architecture} as our network architecture. We
train separate networks for different trajectory lengths, but use the same
architecture for each. Regardless of the trajectory length, all networks
are trained on a total of $10^6$ trajectories from all 5 models. As stated
above, for regression, the data set is equally distributed with respect to the
anomalous diffusion exponents but not among ground truth models, and vice versa
for classification.  Later we also train networks on data sets consisting of
only a single anomalous diffusion model and only $3\times 10^5$ trajectories.
The neural network hyper-parameters, {consisting of learning rate, weight
decay {\cite{krogh_weightdecay}}, batch size, training length (epoch number),
and SWAG moment update frequency}, are tuned using a separate validation set of
$10^4$ trajectories, and final performance results are obtained from a third
testing data set varying in size between $4\times 10^4$ and $1\times 10^5$,
depending on the task. Data are generated using the \texttt{andi-datasets}
python package \cite{andidataset_python}, shorter trajectories are obtained
from the same data set by discarding later data points. Noise, as specified
in equation (\ref{eq_noise}), is added after cutting off the data points
beyond the desired length, as otherwise the signal to noise ratio (snr) on
the long trajectories may not represent the snr of the shortened trajectories,
especially when dealing with models using a changing diffusivity like SBM.

Before training, the trajectory data sets consisting of time series of
positions $x_t$ are pre-processed by conversion to increments $\Delta x_t
= x_{t+1}-x_{t}$ and normalising these increments to a standard deviation
of unity for each trajectory. Rescaling the data in this manner speeds up
the training process and, since we are not interested in a prediction of the
diffusion coefficient, which would be altered by this step, it will not
hinder the neural network's performance.

The networks are trained using the \textit{Adam} optimiser \cite{kingma_adam}
for 65 to 85 epochs with the last 10 to 15 epochs used for SWAG training, where
one epoch corresponds to one full iteration through the training set. The
exact epoch number as well as the other hyper-parameters are fine-tuned
individually for each task and trajectory length using the validation
data set. Once an optimal set of hyper-parameters is found,
we use them to train 20 SWAG models and choose the 5 best performing networks
for \emph{Multi-SWAG}, as measured by their achieved loss on the validation
set. This choice is necessary as some training processes may
get trapped in suboptimal minima.
To obtain the final output, we sample 10 networks from each SWAG model
for a total of 50 Monte Carlo samples and combine these into a
single output of model probabilities for classification or mean and variance
for regression in accordance to equation \ref{eq_mcsample}.

\section*{Data availability}

The data resulting from applying the model on the test data sets are available 
at \url{https://github.com/hseckler/BDL-for-AnDi}. The training and test data 
sets were randomly generated using the \texttt{andi-datasets} python 
package \cite{andidataset_python}.

\section*{Code availability}

All software used in this study is available at 
\url{https://github.com/hseckler/BDL-for-AnDi}.

\begin{acknowledgments}
We thank the German Science Foundation (DFG, grant no. ME 1535/12-1) for
support.
\end{acknowledgments}

\clearpage

\onecolumngrid

\begin{center}

\textbf{Supplementary information: Bayesian deep learning for error estimation
in the analysis of anomalous diffusion}\\[0.4cm]

Henrik Seckler$^1$ and Ralf Metzler$^{1,2}$\\
\textit{$^1$Institute for Physics \& Astronomy, University of Potsdam, 14476
Potsdam-Golm, Germany}\\
\textit{$^2$ Asia Pacific Centre for Theoretical Physics, Pohang 37673,
Republic of Korea}\\[0.8cm]

\end{center}

\twocolumngrid

\renewcommand\thesection{S\arabic{section}}
\setcounter{section}{0}
\setcounter{equation}{0}
\setcounter{figure}{0}
\renewcommand\thefigure{S\arabic{figure}}

\section{Supplementary Method 1}
\label{sec_methods}

For the regression of the anomalous diffusion exponent $\alpha$ we evaluate
the mean absolute error (MAE), defined as
\begin{equation}
\text{MAE}=\frac{1}{N}\sum_{i=1}^N|\alpha_{i,\text{gt}}-\alpha_{i,\mathrm{
pred}}|,
\end{equation}
where $\alpha_{i,\text{gt}}$ and $\alpha_{i,\text{pred}}$ are the ground truth
and predicted anomalous diffusion exponent of the $i$th of the $N$ trajectories
contained in the test data set.

To quantify classification performance we use the accuracy, which is the
fraction of correct classifications.

\paragraph*{\textbf{Reliability diagram and calibration error}}
In order to assess the quality of uncertainty predictions we use reliability
diagrams \cite{guo_reliability}. In these one illustrates observed errors
as a function of predicted uncertainties. For classification tasks we divide
the interval $[0,1]$ into $M$ bins $I_m=(\frac{m-1}{M},\frac{m}{M}]$. If $B_m$
is the set of trajectories with predicted confidences $p_i\in I_m$, then
the accuracy in this interval is given as
\begin{equation}
\text{acc}(B_m)=\frac{1}{|B_m|}\sum_{i\in B_m}\mathbf{1}(\hat{y}_i=y_i),
\end{equation}
where $\hat{y}_i$,$y_i$ are the ground truth and predicted model of input $i$.
The mean predicted confidence of this set is
\begin{equation}
\text{conf}(B_m)=\frac{1}{|B_m|}\sum_{i\in B_m}p_i,
\end{equation}
where $p_i$ is the predicted confidence of the model prediction of the
$i$th input. In a perfectly calibrated model accuracy and confidence
coincide for all bins, which corresponds to the diagonal in the reliability
diagram. Any deviations from the identity represent miscalibration, which
we summarise using the \textit{expected calibration error} (ECE), defined
as \cite{guo_reliability,naeini_ece}
\begin{equation}
\mathrm{ECE}=\sum_{m=1}^M\frac{|B_m|}{N}\left|\mathrm{acc}(B_m)-\mathrm{
conf}(B_m)\right|,
\end{equation}
where $N$ is the number of samples in the test set.

Similarly, we can construct a reliability diagram for the regression
of the anomalous diffusion exponent \cite{levi_ence}. Here the roles of the
mean confidence and the accuracy are taken by the predicted root mean
variance (RMV) and the observed root mean squared error (RMSE). By
introducing a binning of the predicted standard deviation into intervals
$I_m=((m-1)\Delta_\sigma,m\Delta_\sigma]$ of size $\Delta_\sigma$, we
define $B_m$ as the set of trajectories with a predicted standard deviation
$\sigma_{i,\text{pred}}\in I_m$ and obtain
\begin{eqnarray}
\mathrm{RMSE}(B_m)&=&\sqrt{\frac{1}{|B_m|}\sum_{i\in B_m}(\alpha_{i,\mathrm{
gt}}-\alpha_{i,\mathrm{pred}})^2}\\
\mathrm{RMV}(B_m) &=& \sqrt{\frac{1}{|B_m|}\sum_{i\in B_m}(\sigma_{i,\mathrm{
pred}})^2}.
\end{eqnarray}
As above, coinciding RMSE and RMV in all bins represent a perfectly calibrated
model. Deviations from the ideal error prediction are represented by the
\textit{expected calibration error} (ECE)
\begin{equation}
\mathrm{ECE}=\sum_{B_m}\frac{|B_m|}{N}|\mathrm{RMV}(B_m)-\mathrm{RMSE}(B_m)|,
\end{equation}
which can be modified to obtain the \textit{expected normalised calibration
error} (ENCE) \cite{levi_ence}
\begin{equation}
\mathrm{ENCE}=\sum_{B_m}\frac{|B_m|}{N}\frac{|\mathrm{RMV}(B_m)-\mathrm{
RMSE}(B_m)|}{\mathrm{RMV}(B_m)}.
\end{equation}

\section{Supplementary Figures}

Here we provide additional figures with details referenced in the main text.

\begin{figure*}
\includegraphics[width=0.48\linewidth]{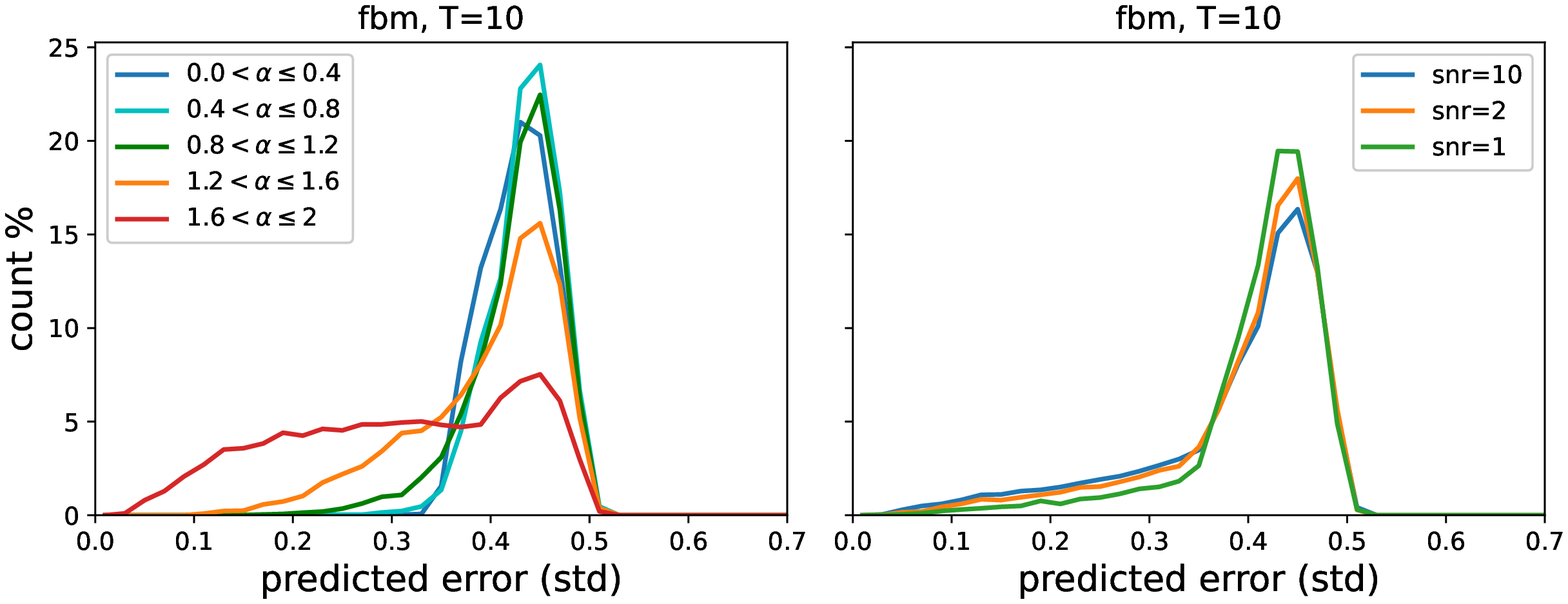}
\includegraphics[width=0.48\linewidth]{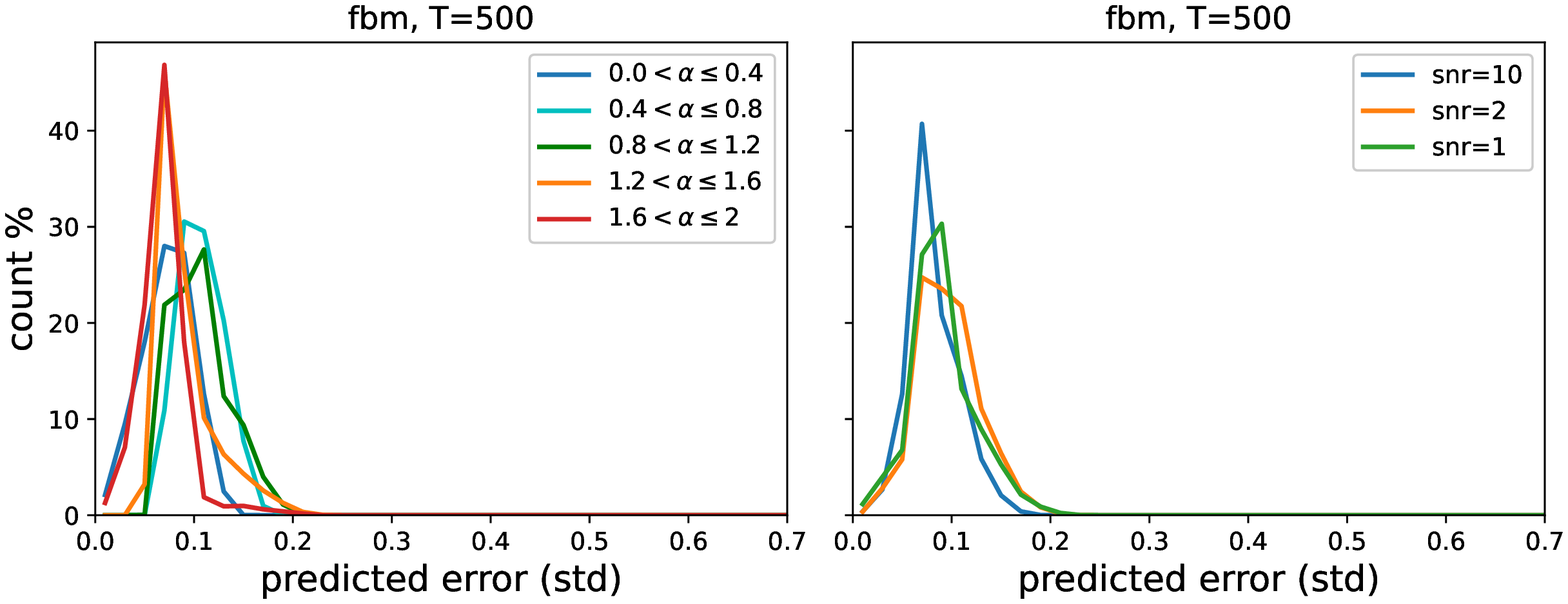}
\includegraphics[width=0.48\linewidth]{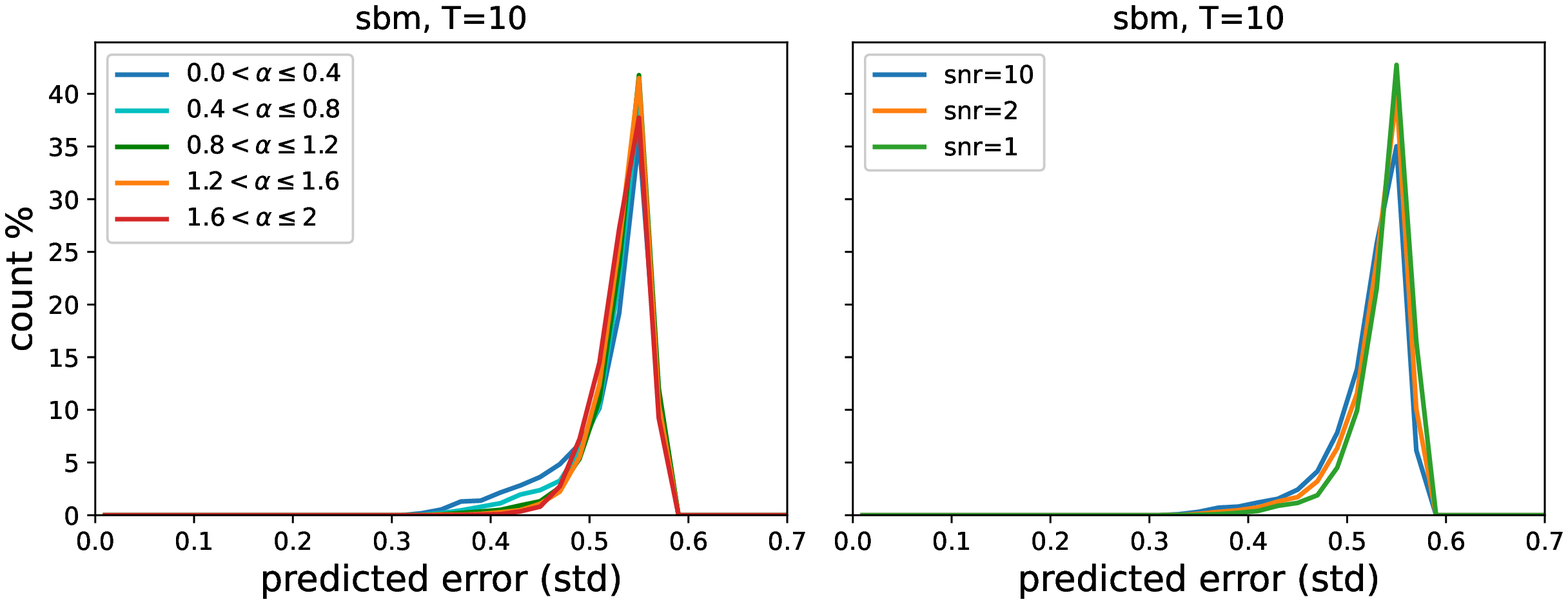}
\includegraphics[width=0.48\linewidth]{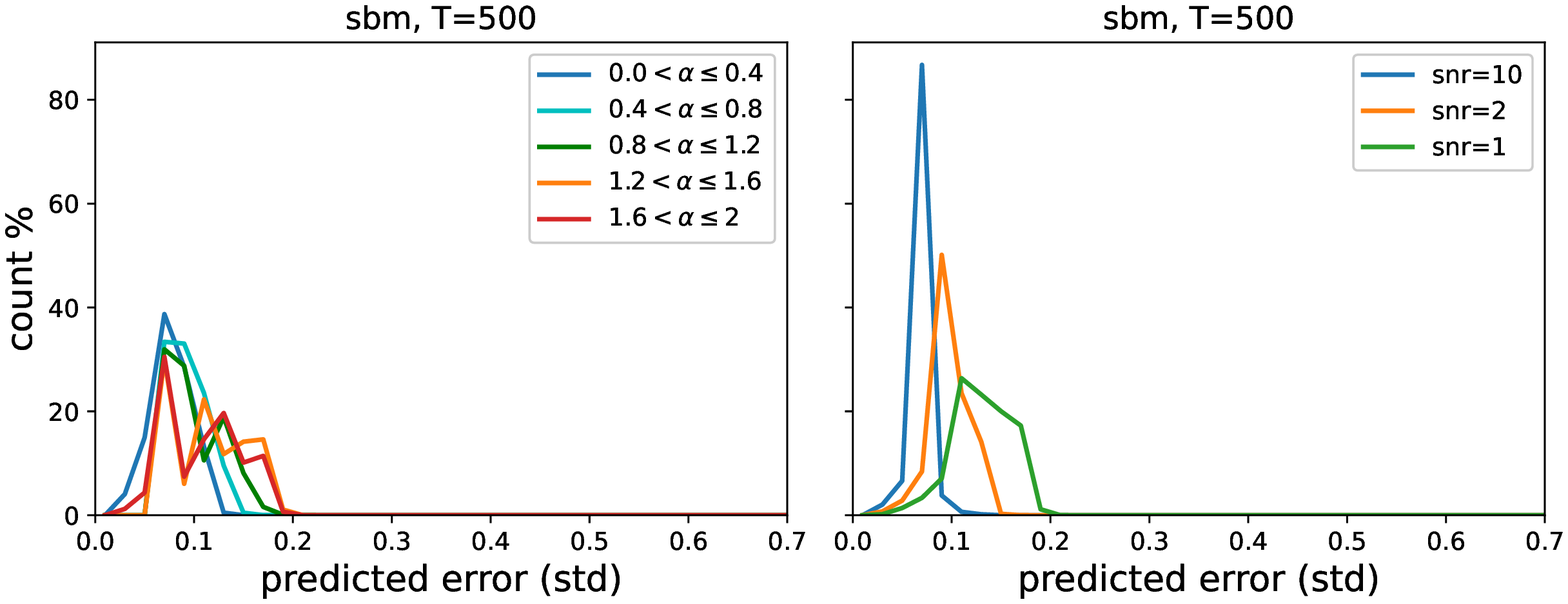}
\includegraphics[width=0.48\linewidth]{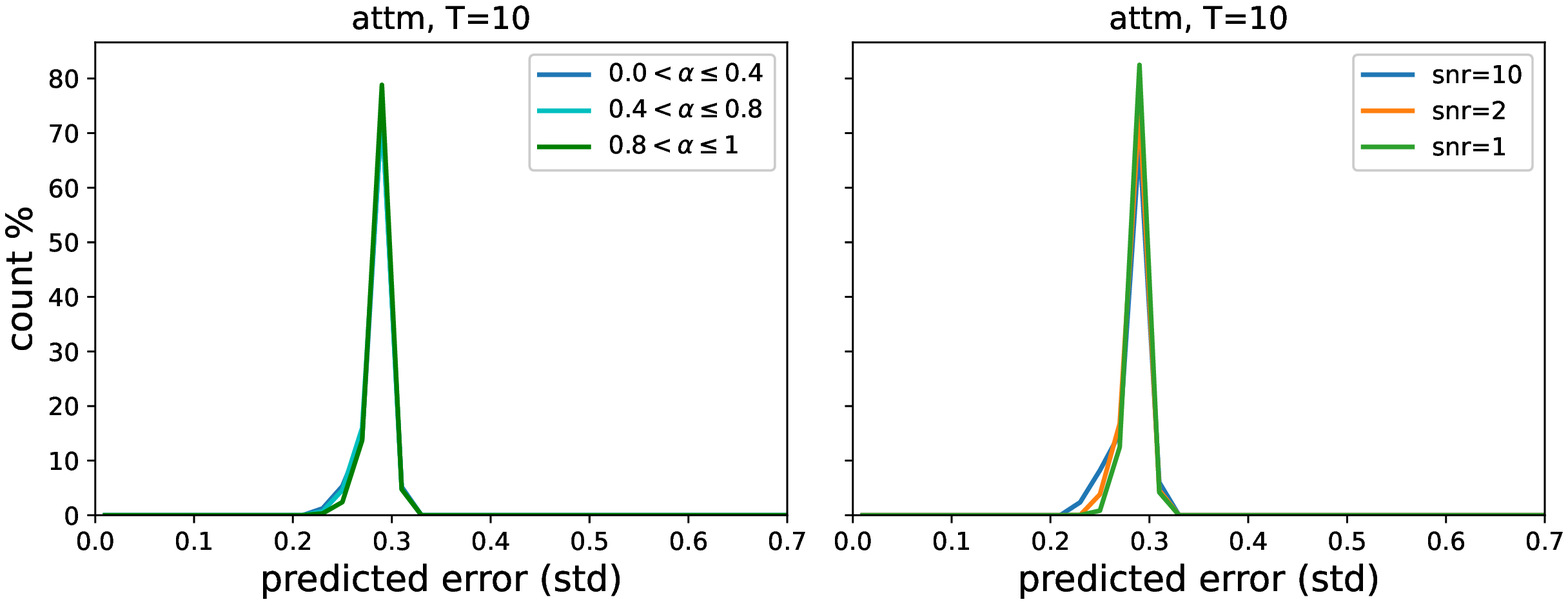}
\includegraphics[width=0.48\linewidth]{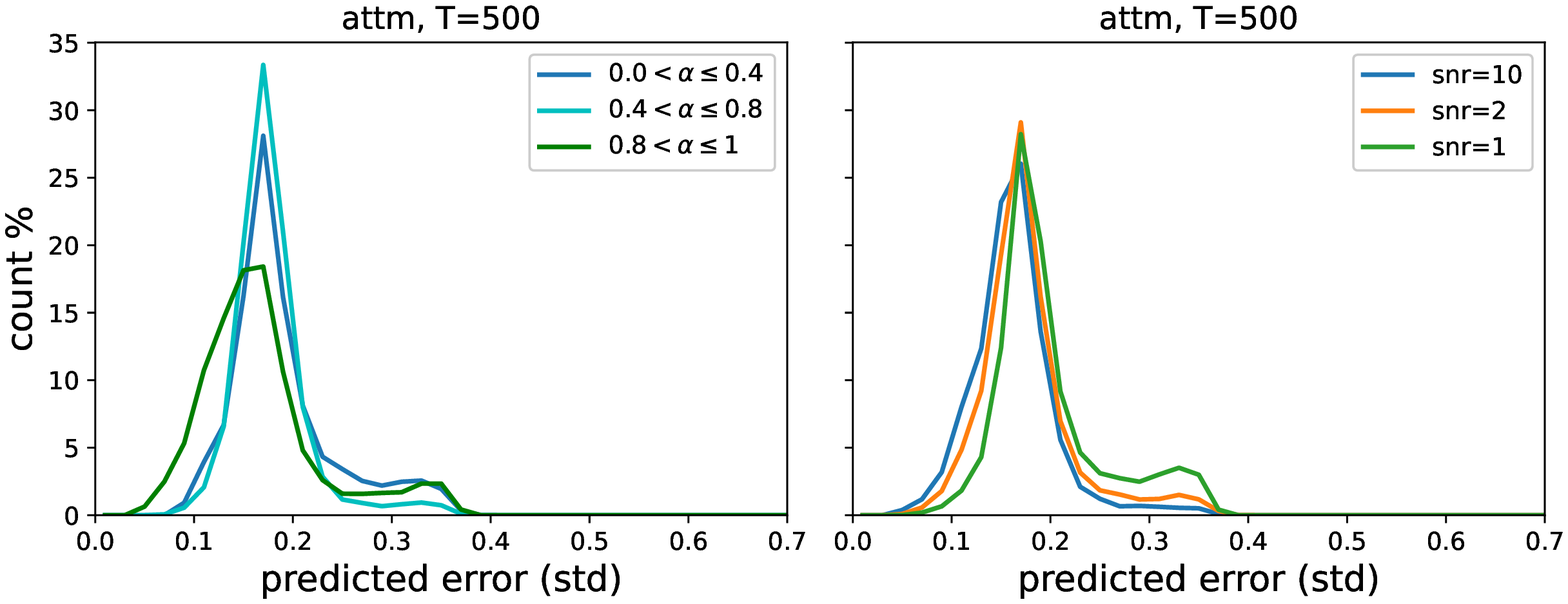}
\includegraphics[width=0.48\linewidth]{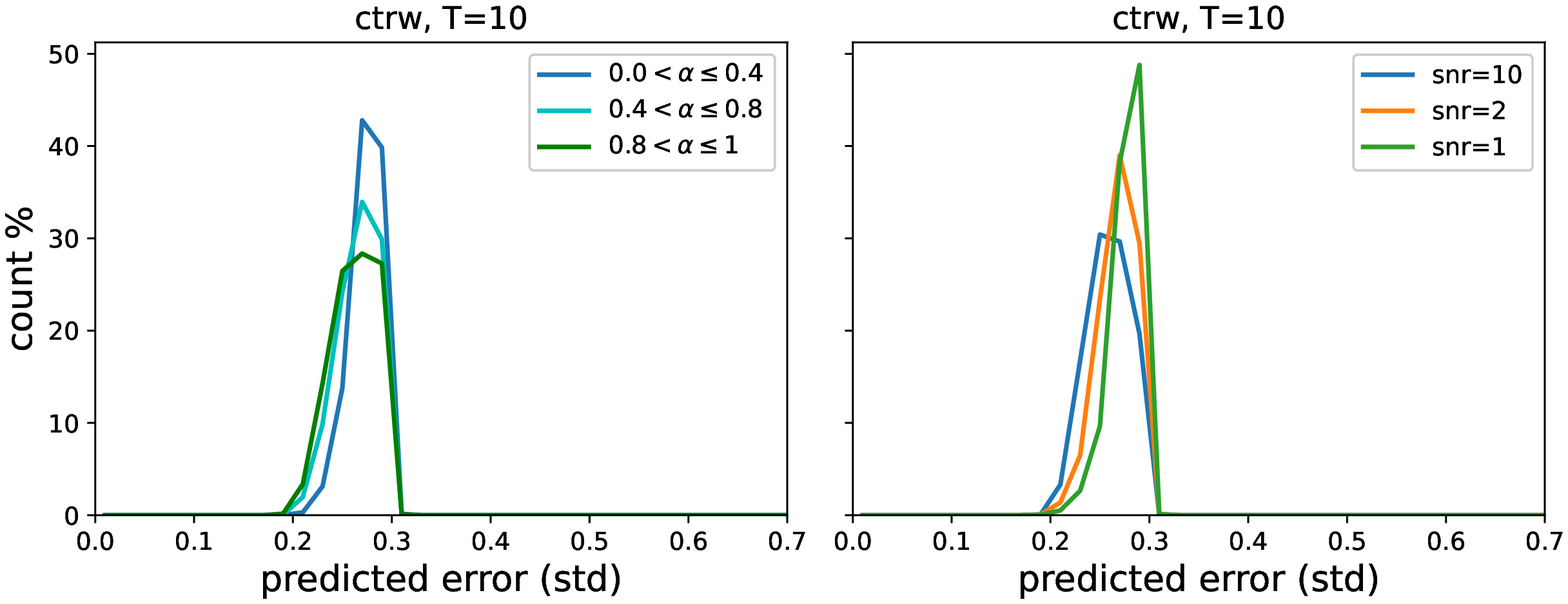}
\includegraphics[width=0.48\linewidth]{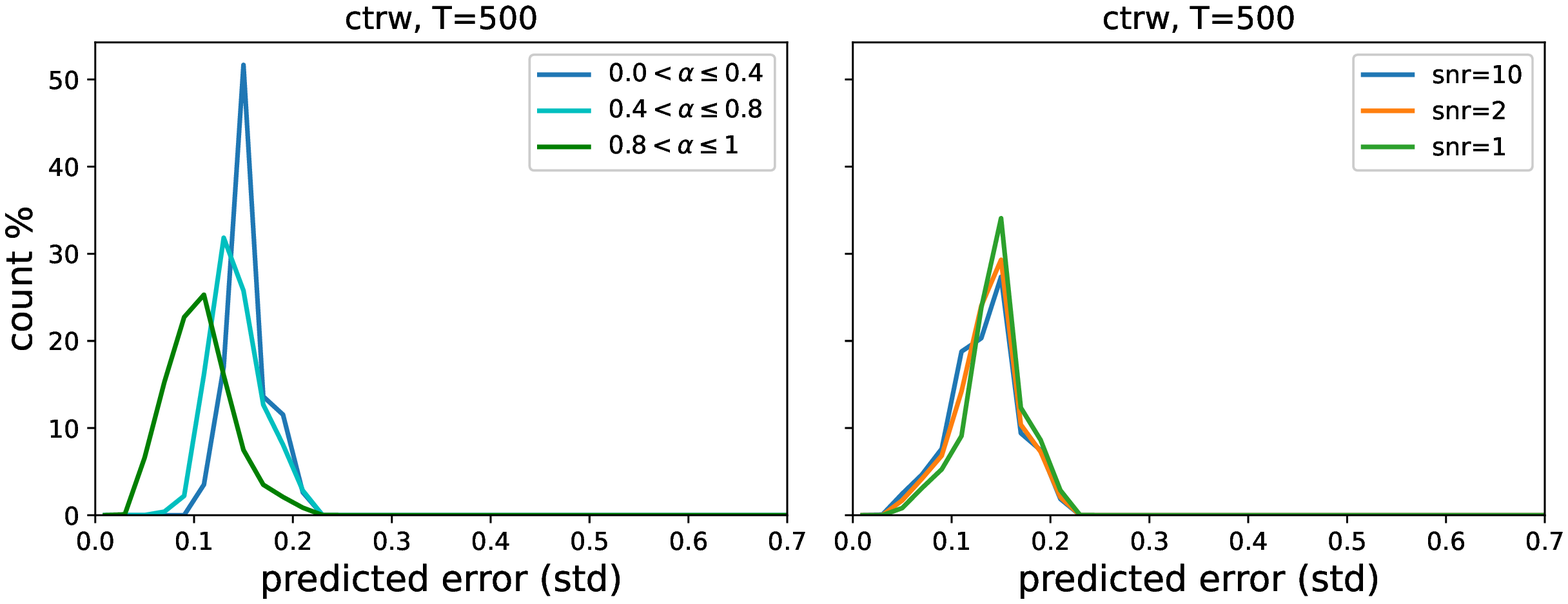}
\includegraphics[width=0.48\linewidth]{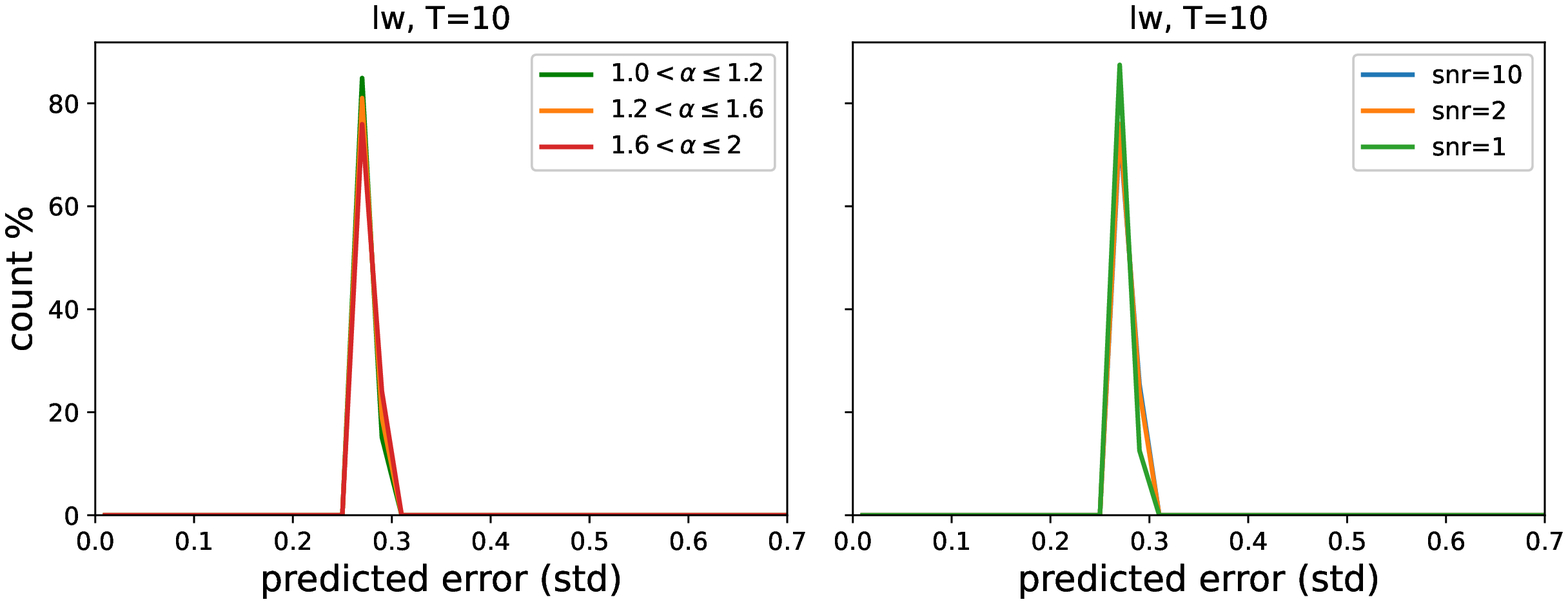}
\includegraphics[width=0.48\linewidth]{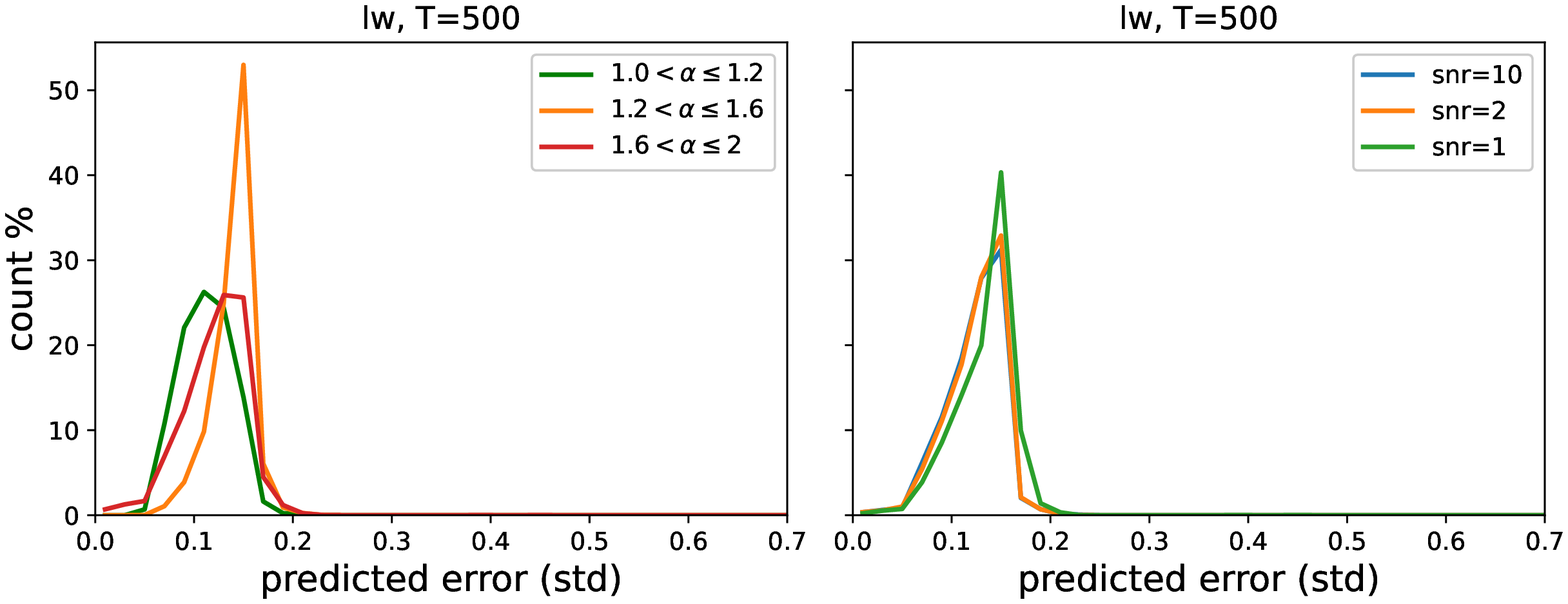}
\caption{\label{fig_histo_otherlenghts} Predicted error histograms split by
exponent and noise for each model for lengths $T=10$ and $T=500$. The used
networks were trained on data sets only containing the one respective diffusion
model and the results are obtained from predictions based on $5\times 10^{4}$
(FBM,SBM) or $4\times 10^{4}$ (ATTM,LW,CTRW) trajectories.}
\end{figure*}

\begin{figure*}
(a) \includegraphics[width=0.95\linewidth]{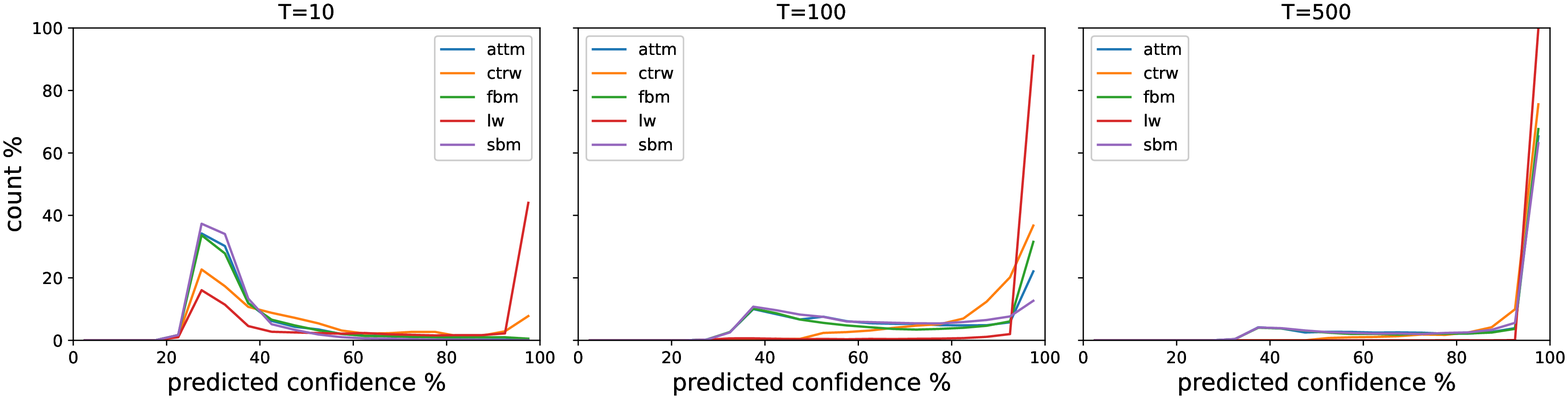}\\
(b) \includegraphics[width=0.95\linewidth]{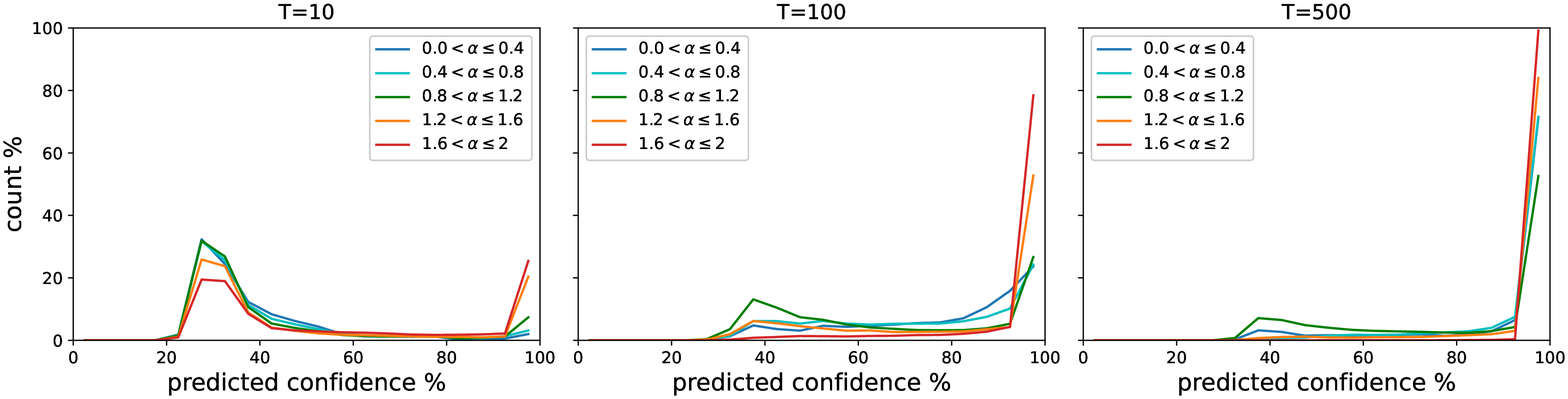}\\
(c) \includegraphics[width=0.95\linewidth]{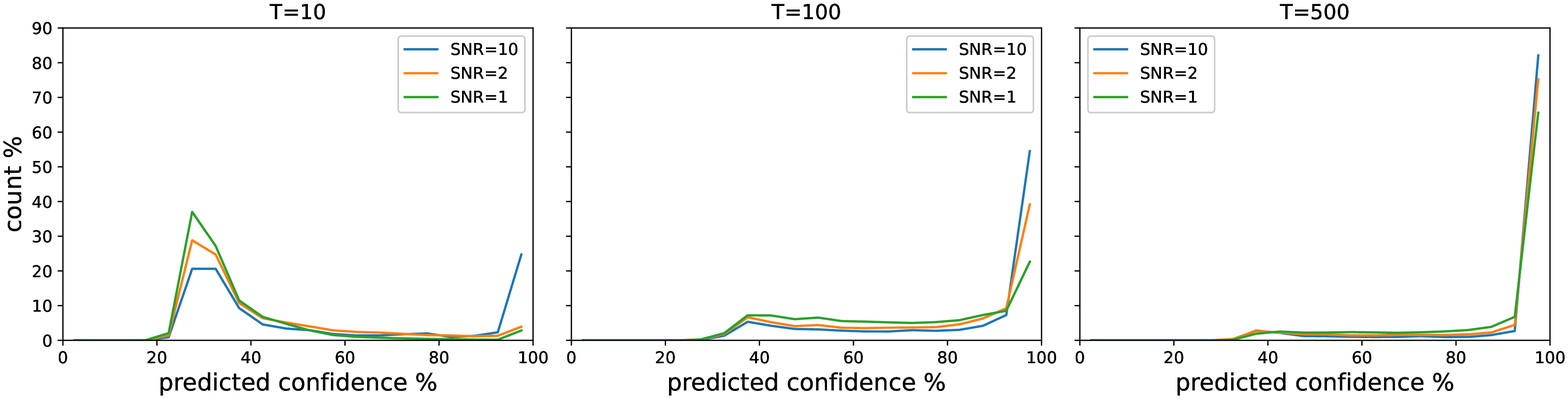}
\caption{\label{fig_class_hist} Error histograms for classification split by
(a) ground truth model, (b) ground truth exponent, and (c) by the used noise.}
\end{figure*}

\end{document}